\documentclass[traditabstract]{aa}  
\pdfoutput=1
\usepackage{graphicx}
\usepackage{txfonts}
\usepackage{natbib}
\usepackage{color}
\usepackage{gensymb}
\def\apj{ApJ}
\def\apjl{ApJ Letters}
\def\aap{A \& A}

\def\mnras{MNRAS}
\def \actaa{Acta Astronomica}

\def \araa{ARA\&A}

\newcommand{\Ared}{\color{black}}

\begin{document}
\title{Statistics of Stellar Variability from \emph{Kepler} - I: Revisiting Quarter 1 with an Astrophysically Robust Systematics Correction}
\author{A. McQuillan\inst{1}\fnmsep\thanks{\email{Amy.McQuillan@astro.ox.ac.uk} } \and
  S. Aigrain\inst{1}  \and S. Roberts\inst{2} }
\institute{Oxford Astrophysics, Keble Road, Oxford OX1 3RH, UK   \and  
Oxford Department of Engineering Science, Parks Road, Oxford OX1 3PJ, UK}
\date{Received 15 November 2010 / Accepted 23rd Nov 2011} 

\abstract{We investigate the variability properties of main sequence stars in
the first month of \emph{Kepler} data, using a new astrophysically robust
systematics correction. We find that the fraction of stars with
variability greater than that of the Sun is {\Ared 60\%, which is
marginally consistent with previous studies}, and confirm the trend of increasing
variability with decreasing effective temperatures. We define low and
high variability samples, with a cut corresponding to twice the
variability level of the active Sun, and compare the properties of the
stars belonging to each sample. We show tentative evidence that the
more active stars have lower proper motions and {\Ared may be} located closer
to the galactic plane. We also investigate the frequency content of
the variability, finding clear evidence for periodic or quasi-periodic
behaviour in 16\% of stars, and showing that there
exist significant differences in the nature of variability between
spectral types. Of the periodic objects, most A and F stars have short
periods ($<2$ days) and highly sinusoidal variability, suggestive of
pulsations, whilst G, K and M stars tend to have longer periods ($>5$
days, with a trend towards longer periods at later spectral types) and
show a mixture of periodic and stochastic variability, indicative of
activity. Finally, we use auto-regressive models to characterise the
stochastic component of the variability, and show that its typical
amplitude and time-scale both increase towards later spectral types,
which we interpret as a corresponding increase in the characteristic
size and life-time of active regions.}


\titlerunning{Statistics of Stellar Variability from \emph{Kepler}}
\maketitle
\section{Introduction}  
\label{sec:intro}

The past decade has seen a rapid increase in the rate of discovery and classification of variable stars, mainly as a result of time-domain photometric surveys whose primary goals were to search for other phenomena, such as microlensing or planetary transits (e.g.\ the OGLE survey, \citealt{uda08}). Sources of stellar variability are wide ranging, from the potentially large-amplitude signatures of eclipses, star spots and pulsations, down to the sub-millimagnitude changes induced by granulation. The typical precision and time sampling of ground-based surveys confines the associated variability studies to `classical' variable stars, with amplitudes of a percent or more. In these surveys, the Sun, whose total output never varies by more than around 0.5\% peak-to-peak {\Ared \citep{fro11}}, even at the maximum of its activity cycle, would appear as a `quiet' or `constant' star. However, space-based transit surveys such as CoRoT \citep{bag03} and \emph{Kepler} \citep{bor10} are sensitive to `micro-variability' down to and well below the solar level, on timescales ranging from minutes to months and, over the entire lifetime of \emph{Kepler}, years.

Measuring the basic characteristics of the variability (amplitude, periodicity, etc\ldots) across large samples of stars, and comparing them to stellar parameters such as age, mass and composition, is a first step towards a better understanding of the underlying phenomena. Many of the latter are ill-understood, because they are related to {\Ared rotation,} convection and magnetism which are challenging to model.  Variability statistics also have a crucial impact on exoplanet studies, particularly for radial-velocity searches or radial-velocity confirmation of transiting planet candidates \citep[see e.g.][]{pon10}.

Data from the \emph{Kepler} mission is particularly amenable to statistical variability studies because of the instrument's unprecedented {\Ared photometric precision\footnote{{\Ared The \emph{Kepler} mission requires photometric precision of 20 ppm on 6.5 hr timescales for a \emph{Kepler} magnitude 12 star, see \cite{jen10} for a detailed discussion.}}} and vast field of view {\Ared (115 deg$^2$)}. The Quarter 1 (Q1) data, which was made public in June 2010, has already been studied by \citet[][hereafter B10]{bas10}, who show that somewhat less than half of the dwarf stars surveyed by \emph{Kepler} are more variable than the Sun on timescales of up to a month, with the fraction increasing from earlier to later spectral types. \citet[][hereafter B11]{bas11} went on to demonstrate that periodic variable stars have significantly larger amplitudes, as a sample, than aperiodic variables. Finally, \citet[][hereafter C11]{cia10} performed a complementary study of the same sample using dispersion rather than amplitude as a variability statistic, and studying likely dwarfs and giants separately using the stellar parameters provided in the \emph{Kepler} Input Catalog (KIC, \citealt{bro11,bat10}). Variability statistics have also been determined using the 10 days of commissioning data (Q0), with the aim of developing methods to characterise and select specific types of variable \citep{wal10}.

In C11, the data were corrected for systematics using the \emph{Kepler} team's Pre-Search Data Conditioning (PDC) method \citep{jen10}. This inspired us to investigate further the apparent bimodality in the variability of dwarf stars observed by \emph{Kepler}, with particular attention to the effect of different systematics correction methods.

The goal of the present paper is to revisit the work of C11 and B10,11 following the application of a new astrophysically robust de-trending method, designed to preserve intrinsic variability signals and remove as fully as possible the systematics. We quantify trends, previously identified and new, between variability characteristics and stellar properties, and investigate the nature of the variability in more detail, in order to gain further insight into the underlying mechanisms.

We describe the data and our systematics correction process in Section~\ref{sec:sysindat}. In Section~\ref{sec:stats_sect} we discuss our choice of variability statistics and examine the fractions and physical properties of the low and high variability samples, focussing on their periodic and stochastic nature in Section~\ref{sec:perstoch}. We present our conclusions and discuss the implications of our findings in Section~\ref{sec:disc}.

\section{Systematics Removal in the Q1 Data} 
\label{sec:sysindat}

\subsection{The Data} 
\label{sec:dat}

The \emph{Kepler} Quarter 1 (Q1) observations took place over $\sim$33.5 days between May 13$^{\rm th}$ and June 15$^{\rm th}$ 2009. 156,097 targets were observed during this time, most with a cadence of 29.42 minutes (a small subset were observed with an increased cadence of $\sim$1 minute, but only the long cadence data was used in this study).   The plate scale is 3.98 arcsec per pixel \citep{vcl09} and the astrometric precision for a single 30 minute measure is better than 4 milliarcsec \citep{mon10}. Known eclipsing binary systems were removed based on the list compiled by \cite{sla11}, available online\footnote{http://keplerebs.villanova.edu/}. \emph{Kepler} planet candidates \citep{bor11} were also removed\footnote{http://archive.stsci.edu/kepler/planet\_candidates.html}. {\Ared Light curves with discontinuities, detected as a jump greater than $4\sigma$ and visually confirmed, were also removed, leaving the final sample size at 123030 stars.}

Both the `raw' and PDC corrected data are publicly available through the \emph{Kepler} mission archive\footnote{http://archive.stsci.edu/kepler} at STScI and also from the NASA Star and Exoplanet Database\footnote{http://nsted.ipac.caltech.edu} (NStED). This study uses the STScI August 2011 data release. The stellar properties used for classification in this study, namely effective temperature (accurate to 200 K) and surface gravity (accurate to 0.5 dex), come from the KIC. These parameters were estimated using Bayesian posterior probability maximisation to match observed colors, estimated from Sloan g, r, i, z filters, {\Ared 2MASS JHK}, and D51 (510nm), to \cite{ck04} stellar atmosphere models. The proper motion values from the KIC are taken from a selection of {\Ared catalogs\footnote{{\Ared \emph{Kepler} Stellar Classification Program, Hipparcos, Tycho-2, UCAC2, 2MASS and USNO-B1.0.}}} where available. Total proper motion is listed on NStED as having accuracy of 20 milliarcseconds per year. 39,000 dwarf stars are listed with non-zero total proper motion. {\Ared The precisions listed in the KIC are the typical value for each parameter, although in reality these may vary by a small amount between stars of different magnitude and spectral type. }  For a more detailed discussion of the KIC parameters, see \cite {bro11,bat10, ver11}.

\subsection{Astrophysically Robust Correction (ARC)} 
\label{sec:arc}

C10 used the PDC corrected flux, whereas B10,11 used the `raw' data, fitting and subtracting a low-order polynomial to remove long-term trends before computing their statistics. 

The PDC correction, which is progressively revised and updated, uses ancillary engineering data such as temperature, pointing and focus variations to remove systematic effects from the light curves. {\Ared The PDC attempts to fit and remove intrinsic variability and then apply the systematics correction to the residuals before re-inserting the stellar signal.  This creates a decision point where the program must determine whether the variability is real or not, which can lead to real low amplitude and long term stellar signal being falsely removed. It can also add high-frequency noise \citep{van10}. The polynomial correction of B10,11 can equally affect real long-term variability so we therefore opted to develop a new astrophysically robust correction (ARC) for systematics.} The ARC will be presented in detail in a forthcoming paper (Roberts et al. in prep.), but we give a brief outline of it here for completeness. 

Our core assumptions are that i) the trends are \emph{systematic} i.e. {\Ared they are instrumental in origin and are therefore present, at some level, in the majority of light curves}, ii) the underlying number of significant trends is unknown and the trend profile is not pre-specified, iii) the amount of trend present changes from light curve to light curve and is unknown and finally iv) although many trends may be curves with a long timescale, there should be no bias towards smoothness.

We consider the observed set of light curves, $d_i$, to be composed of a linear combination of underlying ``true'' curves, $s_i$, an unknown combination of an unknown number, $J$, of systematic trends, $u_j$, and an observation noise process, $\epsilon$,
\begin{equation}
d_i = s_i + \sum_{j=1}^{J} a_{ij} u_j + \epsilon_i
\label{eq: light_curve}
\end{equation}
in which the unknown factors $a_{ij}$ represent the amount of trend $j$ believed to be present in light curve $i$.

To evaluate the number of trends we start by modelling a light curve, $d_n$ say, as a linear combination of all other curves with factor weights $\beta_i$,
\begin{equation}
\hat{d}_n = \sum_{i \not= n} \beta_{i} d_i .
\label{eq:linMod}
\end{equation}
The inference is achieved using a fully Bayesian linear model, based on the computationally efficient variational Bayes methodology and including shrinkage on the $\beta_i$ parameters.

\begin{figure*}
\centerline{(a)\includegraphics[width = 8cm] {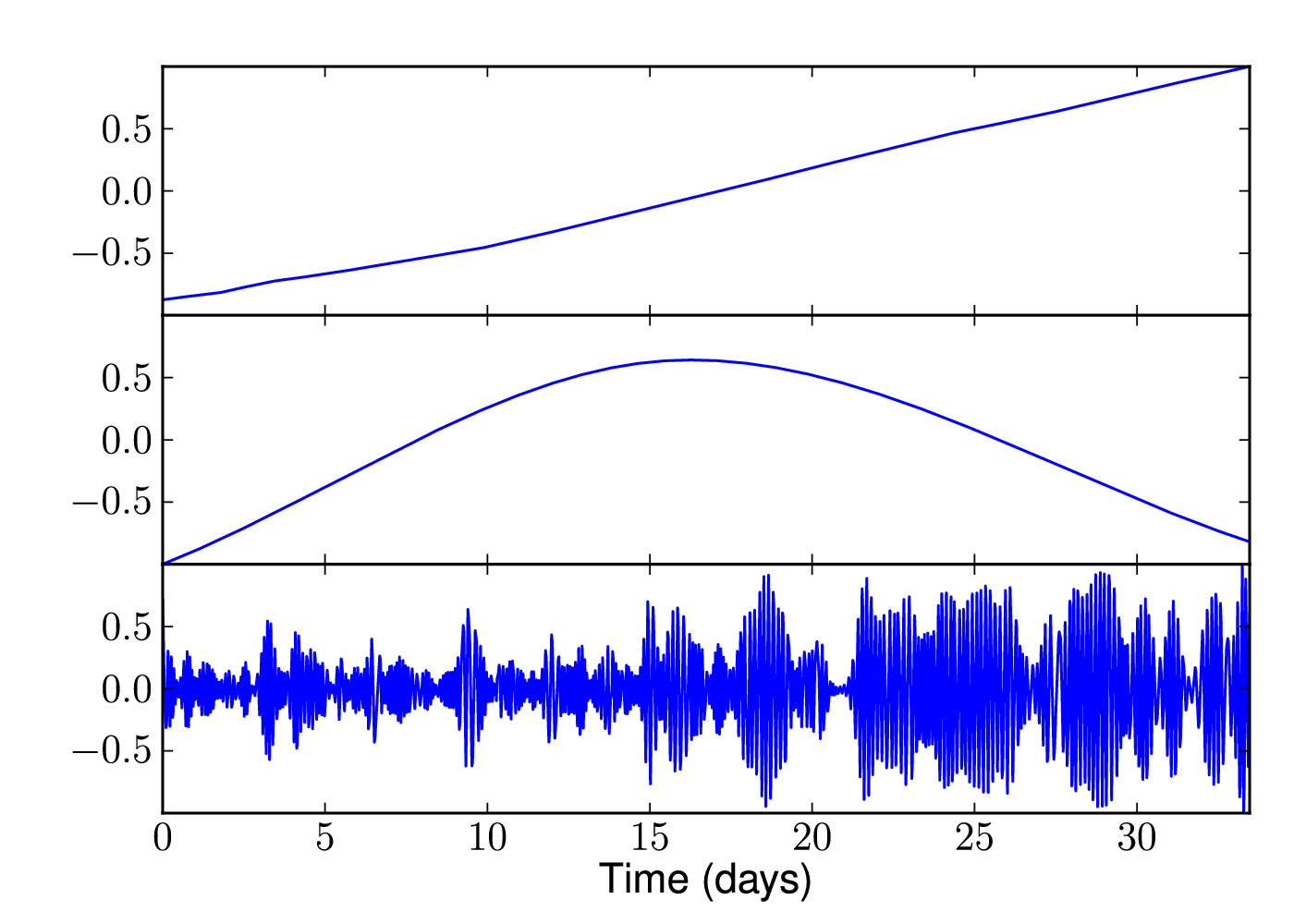}~~
(b)\includegraphics[width= 8cm]{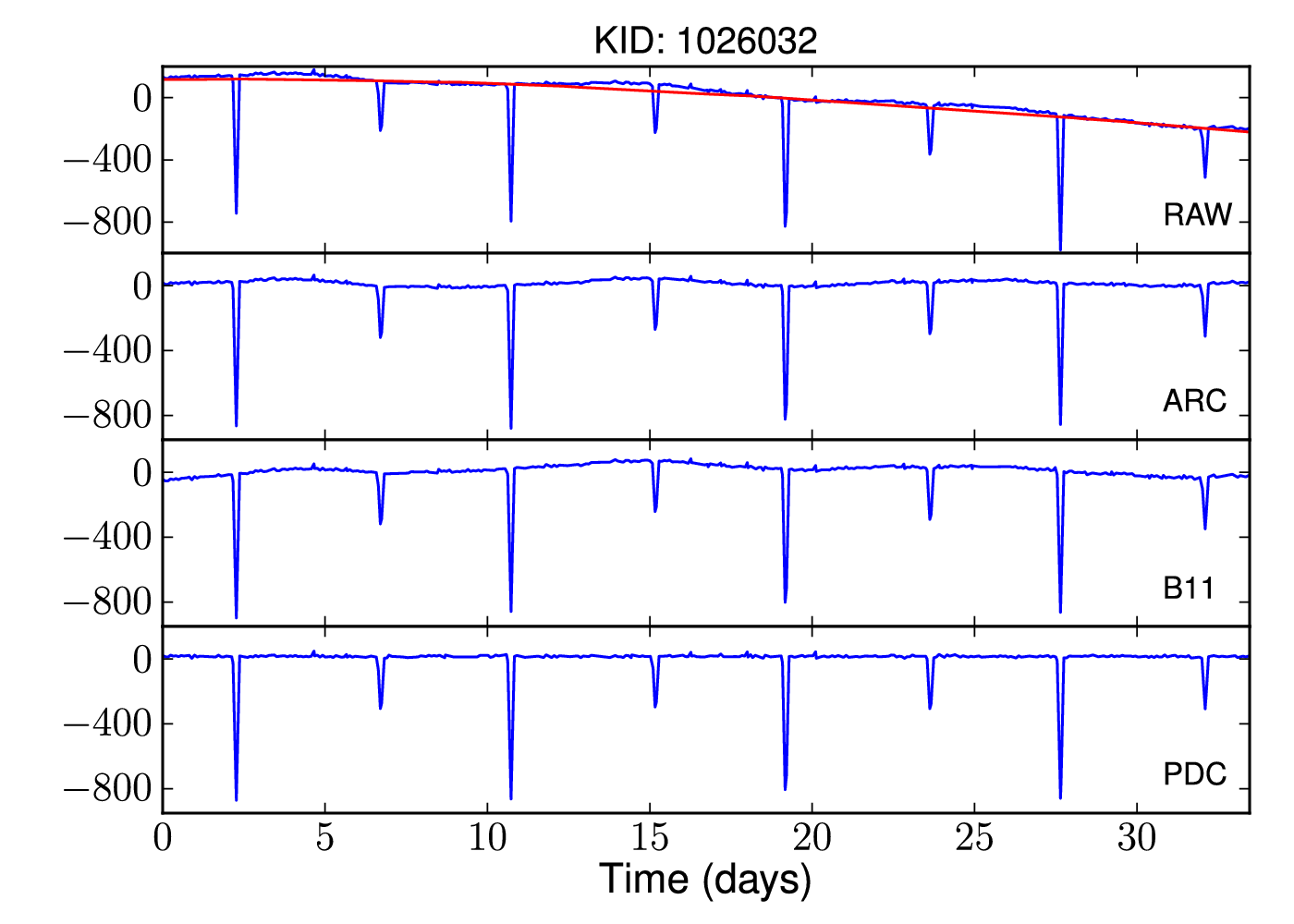}}
\centerline{(c)\includegraphics[width=8cm]{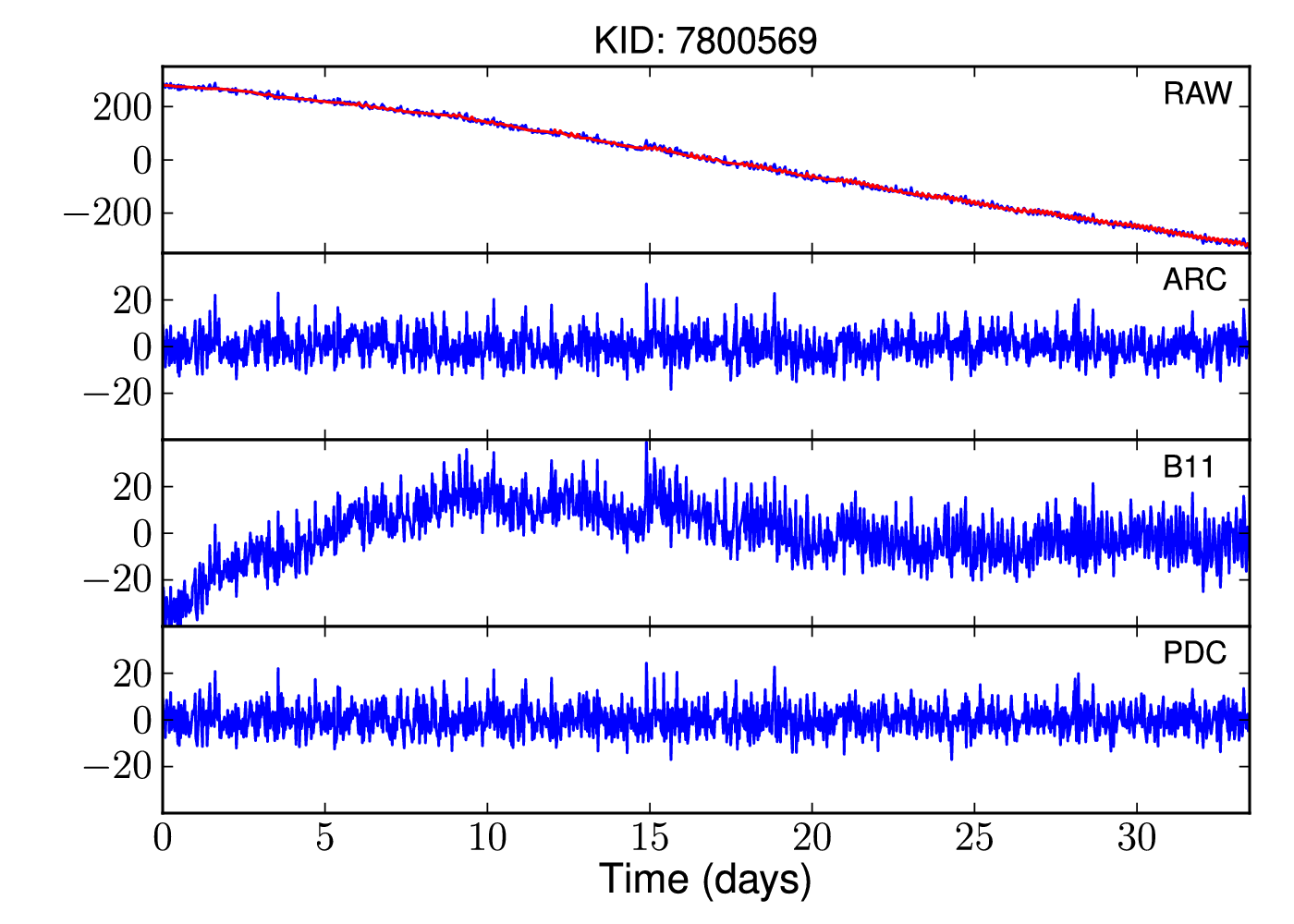}~~
(d)\includegraphics[width= 8cm]{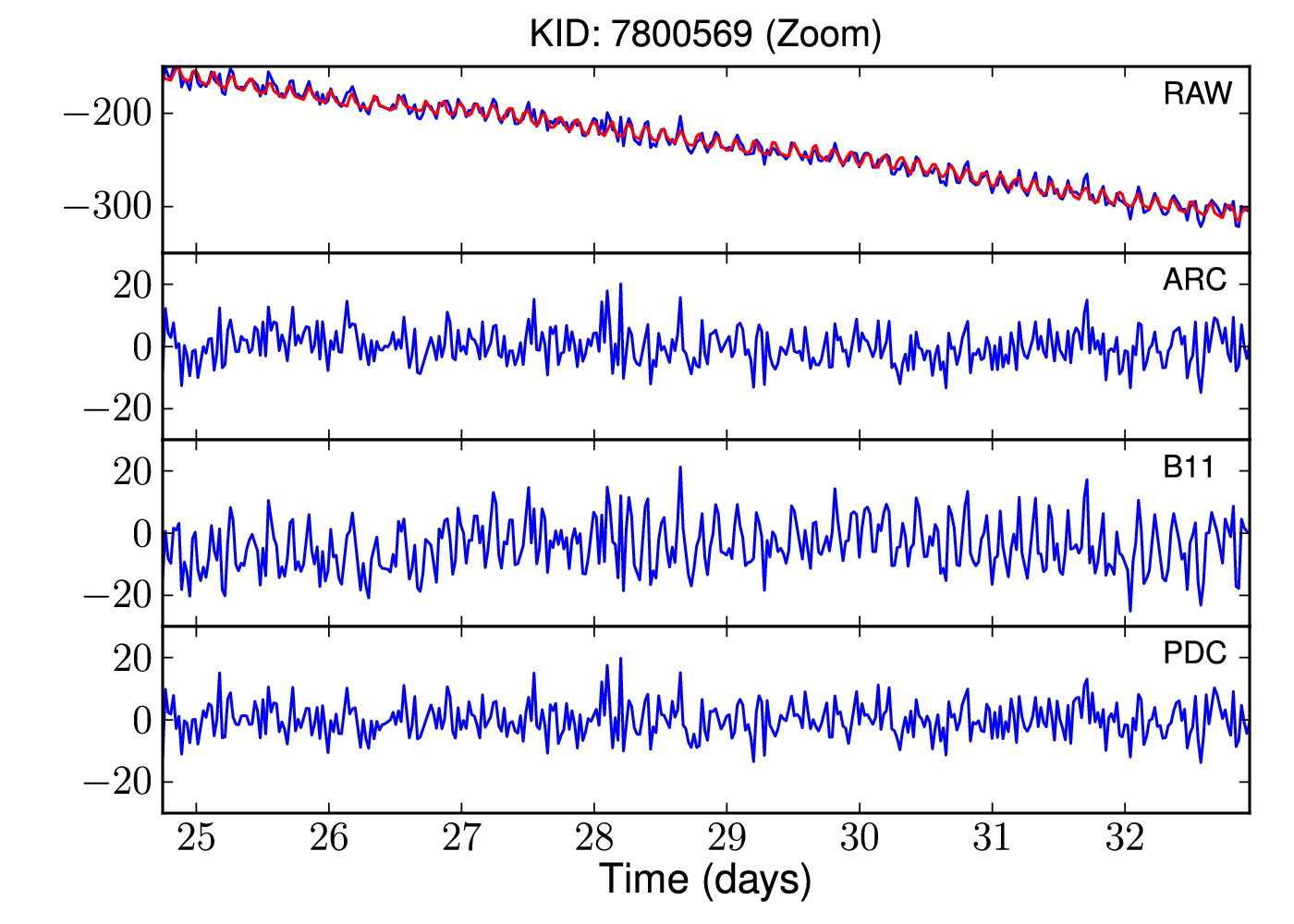}}
\caption{ Subplot {\bf a}) shows example basis trends, inferred from \emph{all} Q1 data; note that the y-axis is in arbitrary units, as these basis functions are scaled to support light curves. Subplots {\bf b}) and {\bf c}) show examples of de-trending and allow comparison between methods: the top trace of each example shows the raw light curve (blue) and the removed trend (red) and the lower subplots show the resultant de-trended light curves using the different correction methods. Subplot {\bf d}) provides detail of a small section of data, highlighting the effective satellite vibration artefact removal obtained.}
\label{fig:eg_results}
\end{figure*}

\begin{figure*}
  \centering
  \includegraphics[width = \linewidth] {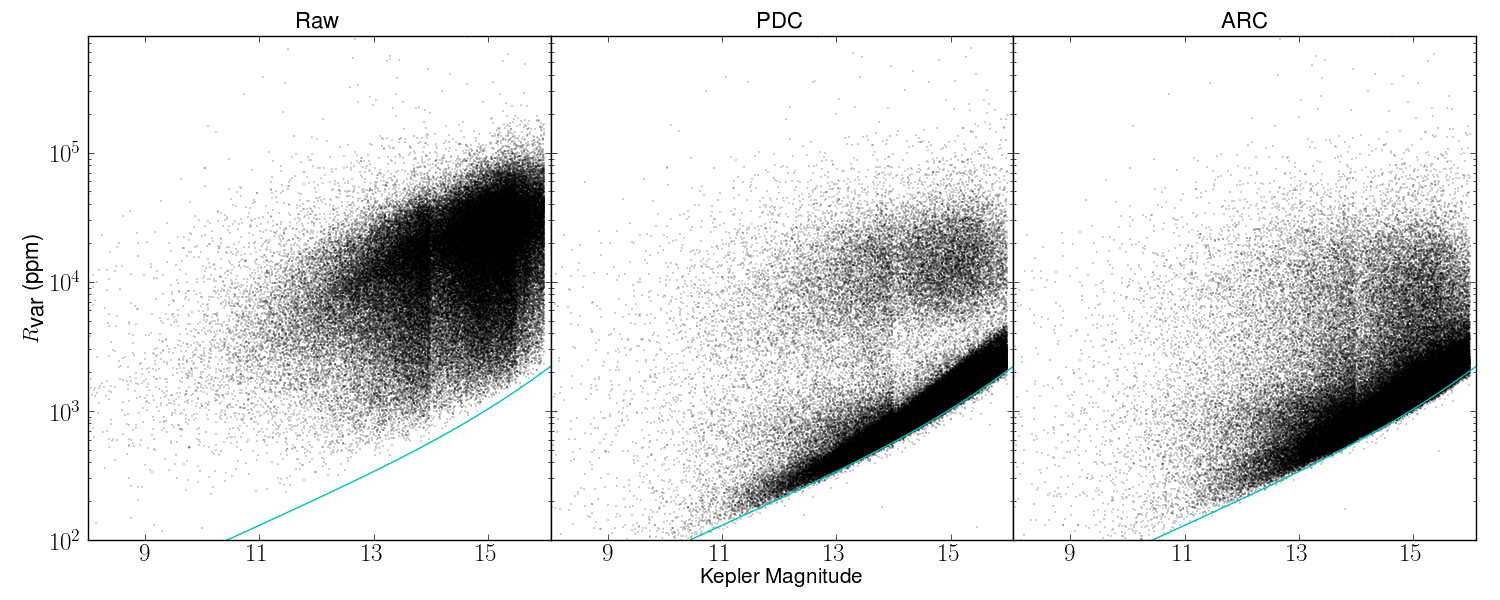}
  \caption{Variability index (described in Section~\ref{sec:var_stats}) against \emph{Kepler} magnitude for the Raw, PDC and ARC \emph{Kepler} Q1 data. The solid line, in the same position on each graph, shows the photometric uncertainty (see Section~\ref{sec:var_frac}). {\Ared The removal of true stellar variability at medium levels by the PDC can be seen in the dearth of stars around $R_{\mbox{var}} = 10^{3} - 10^{4}$ ppm in the middle panel.}}
  \label{fig:comp_var}
\end{figure*}

If we consider the inferred empirical distribution of all $\{ |\beta_i| \}$ we expect systematic trends to be associated with the highest entropy distributions - as true systematic trends are more likely to be present in many light curves rather than in just a small subset. Significant\footnote{We define significance by considering a principal component analysis of the highest entropy trends and require the first principal component to explain more that 95\% of the spectral radius.}
high-entropy trend components are further processed to remove any residual noise, using empirical mode decomposition (EMD) \citep{hua98}, allowing us to recover the dominant dynamic of the trend without imposing any pre-defined smoothness constraints. The latter enables us to extract very low-frequency trends as well as, for example, systematic artefacts due to high frequency vibrations of the satellite. After extraction of the dominant dynamics via EMD, the resultant trend is then removed from all the data by inferring the factors $a_{ij}$ in Equation~\ref{eq: light_curve}, once again using fully Bayesian inference, giving rise to a modified data set which is deflated with respect to the trend. The entire process of trend discovery and removal is then repeated using the deflated data set until no more significant trends are discovered.

Figure \ref{fig:eg_results}(a) shows a typical set of inferred basis trends, obtained from all light curves over all CCDs; note the two typical smooth trends along with the third high frequency vibration effect. Plots (b) and (c) show the raw light curves (blue) along with the inferred trend component (red) and below the de-trended light curves {\Ared for the different correction methods compared in this paper}. Plot (d) highlights a typical section of data in which the high frequency systematic is prevalent and well-removed by the method. {\Ared These plots demonstrate that the ARC provides the most consistently suitable corrections, removing the high frequency trend in (d), maintaining the true stellar variability in (b) and dealing with low frequency smooth systematics effectively in (c). Comparatively it can be seen that neither the PDC or B10 corrections can perform well in all these areas.}

We compared results obtained when correcting all light curves together, with that produced by treating each mod.out\footnote{Each module consists of 2 CCDs, which in turn have 2 outputs each.} separately. The mod.out separated batches are most suitable since they use the same CCD and channel, which leads to more closely correlated systematics, for example, the reaction motor effects are more prevalent on some CCDs than others.

The ARC performs well in the vast majority of cases, removing systematic trends without altering intrinsic stellar variability signals. One remaining effect  that is not currently removed by the ARC is that of the variable guide stars, primarily the eclipsing binary. These introduce small variations in some light curves but these are not frequently occurring or similar enough to be detected by the ARC. A method to detect and remove this effect is currently being devised and will be included in future work, but for this study it may be omitted without significantly reducing the quality of the results.

\section{Variability} 
\label{sec:stats_sect}

\subsection{Variability Statistics} 
\label{sec:var_stats}

In order to ensure that we were using the original release of \emph{Kepler} data correctly, we first reproduced the calculations of B11 as exactly as possible, and compared our results to theirs. We found no discrepancies once the different data reduction methods had been taken into account. 

There are a variety of statistics that can be used to quantify variability. The C11 study use the PDC `dispersion' which can be downloaded from (NStED) together with other pre-computed statistics, including the light curve median and reduced $\chi^2$. Dispersion is defined as the 1 sigma rms scatter around the median magnitude of the light curve.

Instead of dispersion, B10,11 measured the light curve `range', which is essentially a measure of the peak-to-peak variation. The effect of high-frequency noise was removed either by smoothing the light curve on 10-hour timescales (B10) or by discarding the upper and lower 5 percentiles (B11). The choice of statistic used to study the variability is somewhat arbitrary and we confirm that this choice does not significantly alter the results. 

We used the empirical three-section cut of C11 to distinguish between likely dwarfs and giants based on surface gravity $\log g$ and effective temperature $T_{\rm eff}$, rather than the simpler $\log g$ cut used by B10. It has since been noted that the KIC contains some misidentifications \citep{koc10}, but since these are not expected to be numerous enough to affect our results, and for ease of comparison to C11, we used the KIC values without modification.  

Based on the method of B11, we have chosen to use the range $R_{\mbox{var}}$, between the 5th and 95th percentile, for the median normalised light curve as our variability statistic. Dispersion and reduced $\chi^2$ provide an appropriate measure of variability that is believed to be primarily stochastic and Gaussian. Pulsations and rotational variability do not meet these criteria and therefore a measurement based on the peak-to-peak variations in the light curve is considered more relevant. Selecting the 5th to 95th percentile range reduces the noise on the peak-to-peak measurement.

\begin{figure}
  \centering
  \includegraphics[width = \linewidth] {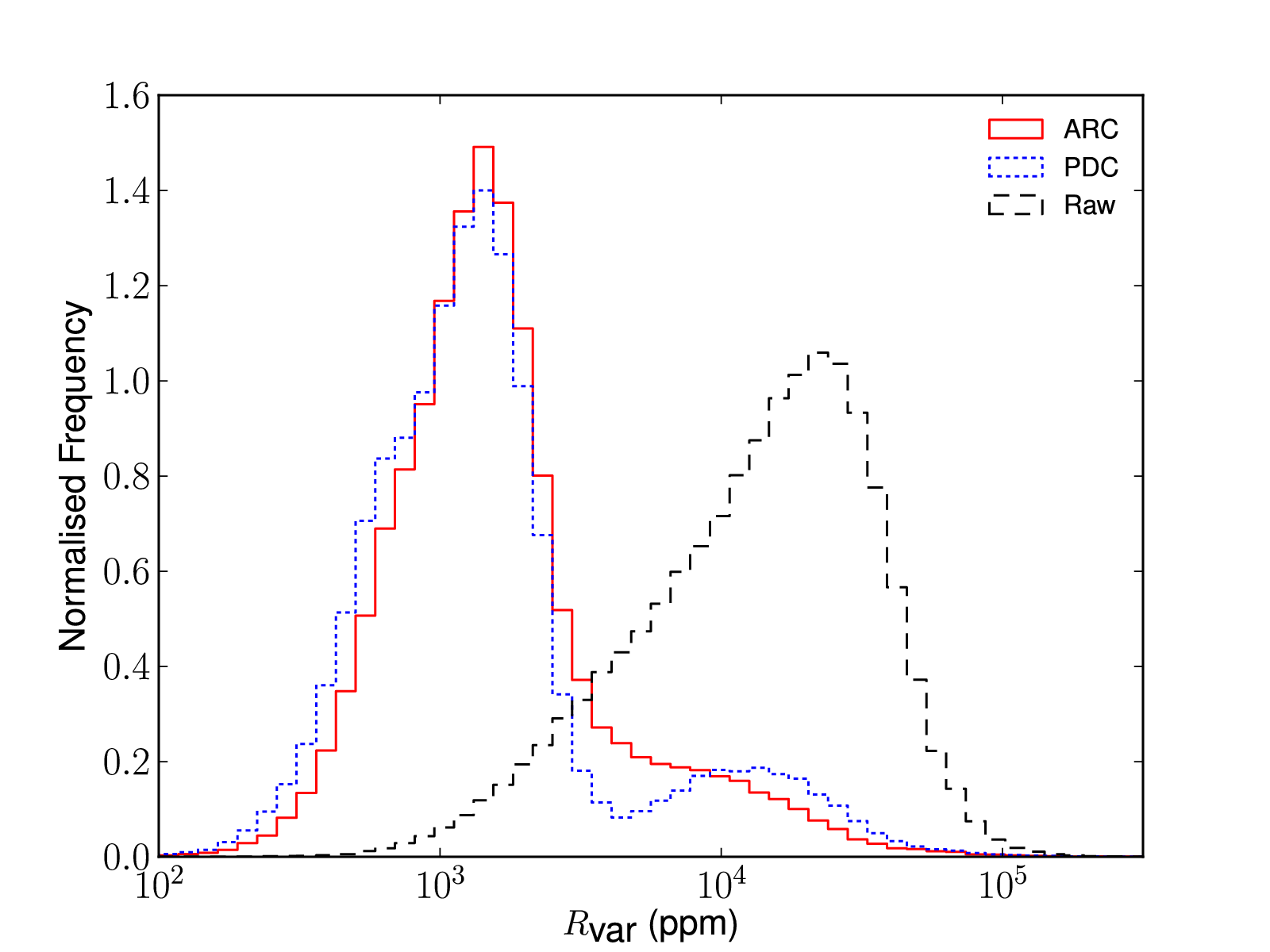}
  \caption{Histogram of variability for stars 13 \textless\ Kepmag \textless\ 14 for the Raw, PDC and ARC \emph{Kepler} Q1 data.}
  \label{fig:comp_var_hist}
\end{figure}

$R_{\mbox{var}}$ measurements for the Raw, PDC and ARC data are compared in Figures~\ref{fig:comp_var} and \ref{fig:comp_var_hist}. The ARC clearly removes most systematic effects, reducing the lower envelope of points to the photo noise limit, however it does not have the side effect of suppressing intermediate amplitude variability (as done by the PDC; this is apparent in the scarcity of points around $R_{\mbox{var}} = 10^{3} - 10^{4}$ ppm in Figure~\ref{fig:comp_var_hist} and the middle panel of Figure~\ref{fig:comp_var}.) Another unfortunate side effect of the PDC is the introduction of high frequency noise in some light curves, which does not occur with the ARC.

\subsection{Solar Comparison and Variability Fraction} 
\label{sec:var_frac}

\begin{figure} 
  \centering
  \includegraphics[width = \linewidth] {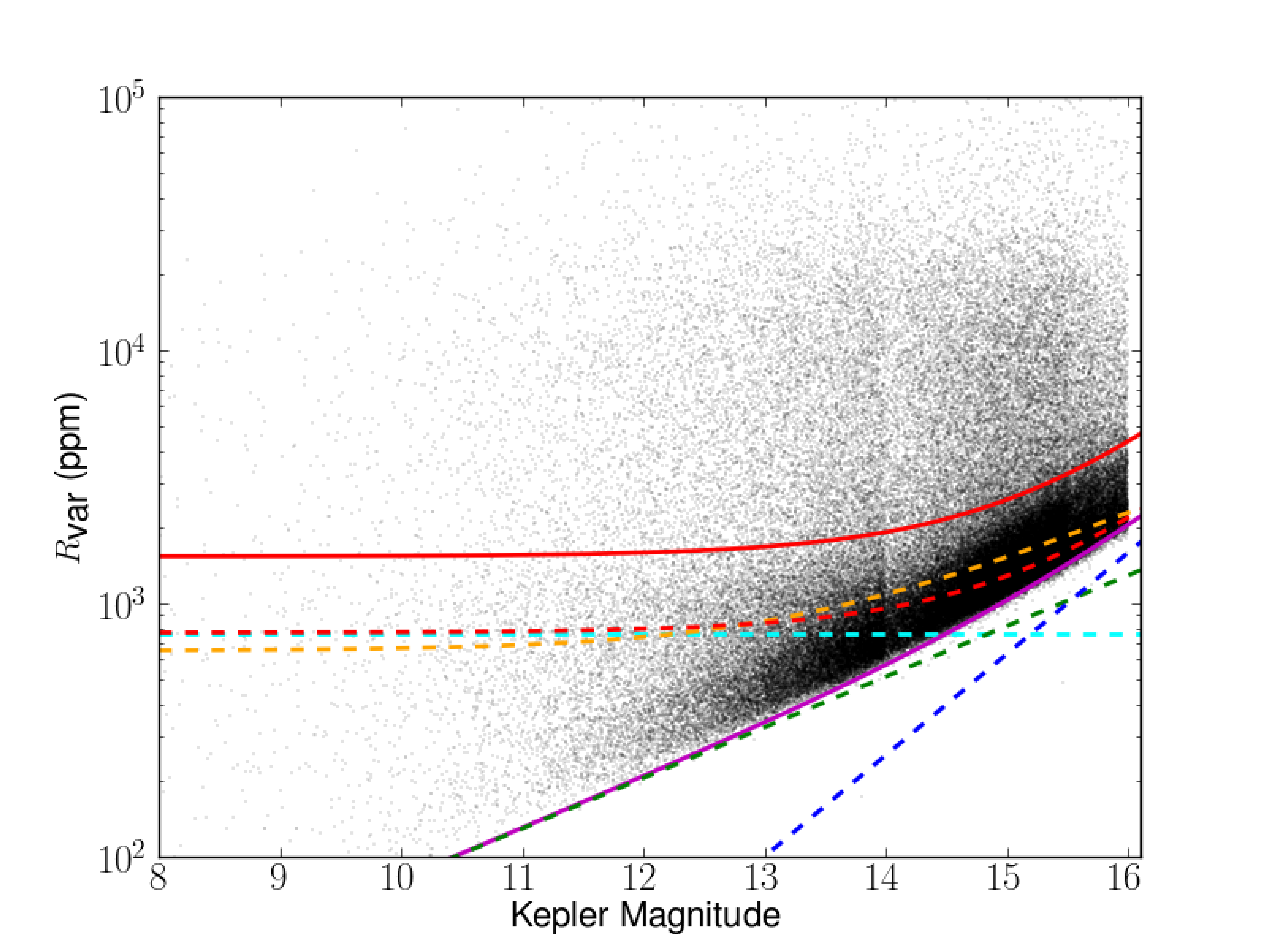}
  \caption{Division of high and low variability dwarf stars (red solid line) at twice the solar value (red dashed line). The solar level is calculated from the active Sun level (cyan dashed line) and the noise (magenta solid line), which itself is a combination of background (blue dashed) and photon noise (green dashed). {\Ared The orange dashed line marks the position of the solar line as determined by B10.}}
  \label{fig:split}
\end{figure}

To compare the properties of the high and low variability stars we make a cut in $R_{\mbox{var}}$ based on a comparison to twice the variability level of the active Sun. {\Ared This level is somewhat arbitrary but provides an appropriate way to divide stars with approximately solar levels of variability and quieter, from those which are significantly more variable than the active Sun.} The solar $R_{\mbox{var}}$ value was calculated from the SOHO/VIRGO summed g+r light curves for the active Sun in the year 2000, because these provide the closest match to the \emph{Kepler} bandpass (B10).  The solar $R_{\mbox{var}}$ value was calculated to be 766 ppm from the average obtained using a sliding 33 day section of data over 2 years, centred on the activity maximum.  An empirical fit to the median of the photometric uncertainties on the light curves in 0.5 mag bins provides an estimate of noise levels across the range of magnitudes. The equation of the solar equivalent line is composed of the solar $R_{\mbox{var}}$ measurement, photon and background noise terms,

\begin{equation} \label{solline}
	y^2 = \mbox{solar }R_{\mbox{var}}^{2}   +  10^{0.4{(\mbox{mag} - p_1)}^{2}} + 10^{1.6{(\mbox{mag} - p_2)}^{2}},
\end{equation}
where best fit values for $p_1$ and $p_2$ are 0.4 and 8.0 respectively.  See Figure~\ref{fig:split} for a graphical representation of the division process.  {\Ared The orange dashed line in Figure~\ref{fig:split} represents the solar line as determined by B10, where extensive visual comparison of solar and \emph{Kepler} light curves is used to estimate a solar level.}

{\Ared The fractions of dwarf stars with variability greater than the solar level are shown in Table~\ref{table:var_fract}. We find that 60\% of dwarf stars are more variable than the active Sun on 33 day timescales.  B10 find 46\% of dwarf stars to be more variable than the solar level. C11 define stars as significantly variable if the reduced $\chi^2$ is greater than 10, and find the fraction of dwarf stars meeting this criteria to be 18\%.  

The variability fraction is strongly dependent on the choice and definition of the solar and photometric noise levels. A larger fraction of stars fall into the low variability subset when the division of B10 is used, due to the small difference in the solar value used and more importantly the position of the estimated photometric noise level. Our empirical fit to the median of the photometric uncertainties produces a line that more closely follows the shape of the lower envelope in Figure~\ref{fig:split} than the visual estimate of B10.

We calculated the random uncertainty for each of the variability and periodicity fractions listed in Table~\ref{table:var_fract} by performing 10,000 measurements using random samples of 80\% of the data, and found these to be \textless\ 0.5\% and hence negligible in comparison to difference introduced by the choice of dividing line.}

{\Ared It is important to note that for many stars in the low variability sample, the lower limit of $R_{\mbox{var}}$ is due to the photometric precision of the \emph{Kepler} satellite, shown as the magenta line in Figure~\ref{fig:split}. However, due to the position of the dividing line (solid red line in Figure~\ref{fig:split}) and its dependence on the photometric precision, there are still a large number of stars that are significantly more variable than the photometric noise limit for their magnitude.} There are $\sim$1,500 stars in the low variability sample that are above twice the photometric noise level.

\begin{table*}
  \caption{Fractions of stars more variable than once and twice the solar level, and fractions of periodic stars in the high and low variability groups (see Section~\ref{sec:per} for definition). {\Ared The random uncertainty for each of the measurements is \textless\ 0.5\%, and is negligible in comparison to difference introduced by the choice of dividing line, demonstrated by the 1 and 2 times the solar value comparisons listed in this table (see Section~\ref{sec:var_frac} for further discussion).}}
  \label{table:var_fract}
  \centering
  \begin{tabular}{lccccccc} 
    \hline\hline
    Sample &  $R_{\mbox{var}}$ \textgreater\  Solar  &  $R_{\mbox{var}}$ \textgreater\ 2$\times$Solar & Periodic Fraction  \\
   \hline
    All Dwarfs   &  0.60  &  0.20  & 0.16 \\ 
    A   &   0.53  &   0.42 &  0.23  \\  
    F  &  0.46  &   0.15   &   0.17   \\  
    G  &  0.56  &   0.14 &  0.13  \\  
    K   &  0.79 &   0.33  & 0.20   \\  
    M  &  0.96   &  0.56 &   0.25 \\   
   \hline
 \end{tabular}
\end{table*}

\subsection{Comparison of Low and High Variability Stellar Properties} 
\label{sec:stelprop}

\begin{figure} 
  \centering
  \includegraphics[width = \linewidth] {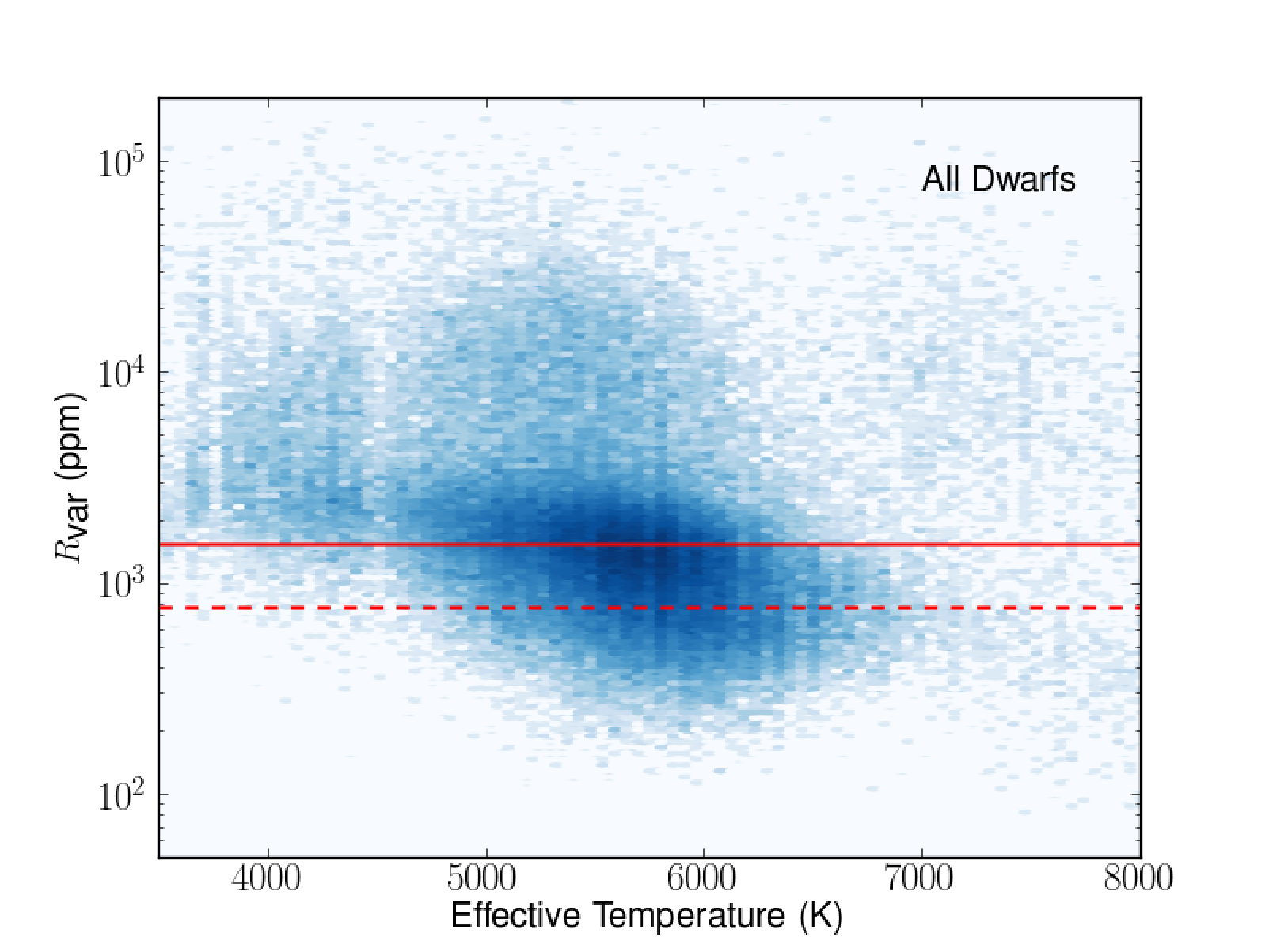}
  \includegraphics[width = \linewidth] {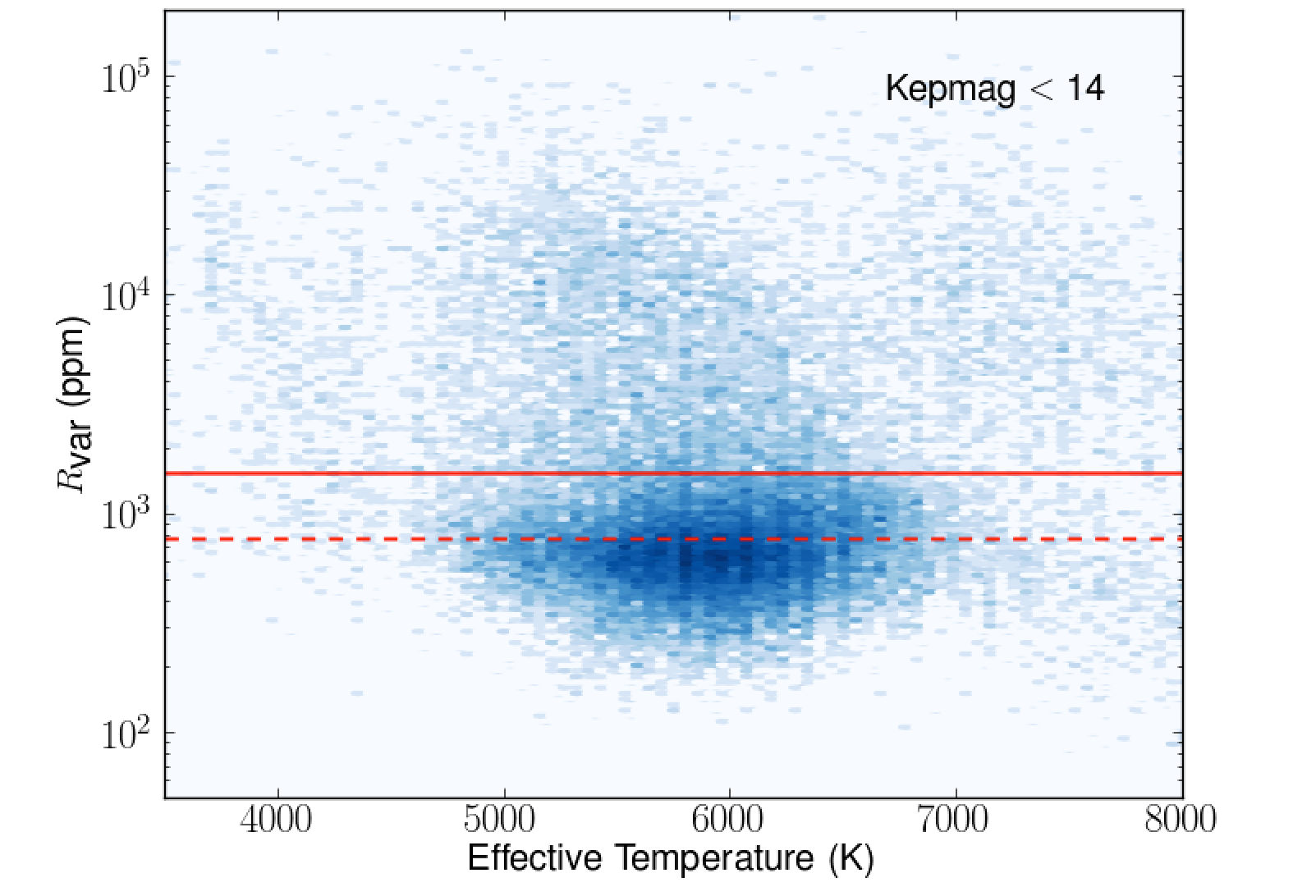}	
  \caption{{\Ared Density plot of effective temperature against $R_{\mbox{var}}$, showing the decrease in variability with increasing temperature for all the selected dwarf stars (\emph{top}) and for those with \emph{Kepler} magnitude \textless\ 14 (\emph{bottom}). This illustrates that the dearth of low variability starts at low temperatures is not a result of the increased noise floor of the cool, faint stars.} The dashed line shows the solar variability level and the solid line is twice the solar level.}
  \label{fig:temp_plot}
\end{figure}

B10,11 and C11 highlighted various relationships between variability and stellar properties. We revisit these using the ARC data.  As C11 noted, there is a decrease in variability with temperature which cannot be entirely explained by the increased noise levels in the fainter cool stars (see Figure~\ref{fig:temp_plot}). Pearson's correlation coefficient  of effective temperature and ${\mbox{log}}_{10}(R_{\mbox{var}})$ for the whole dwarf sample is -0.31.

\begin{figure*} 
  \centering
  \includegraphics[width = \linewidth] {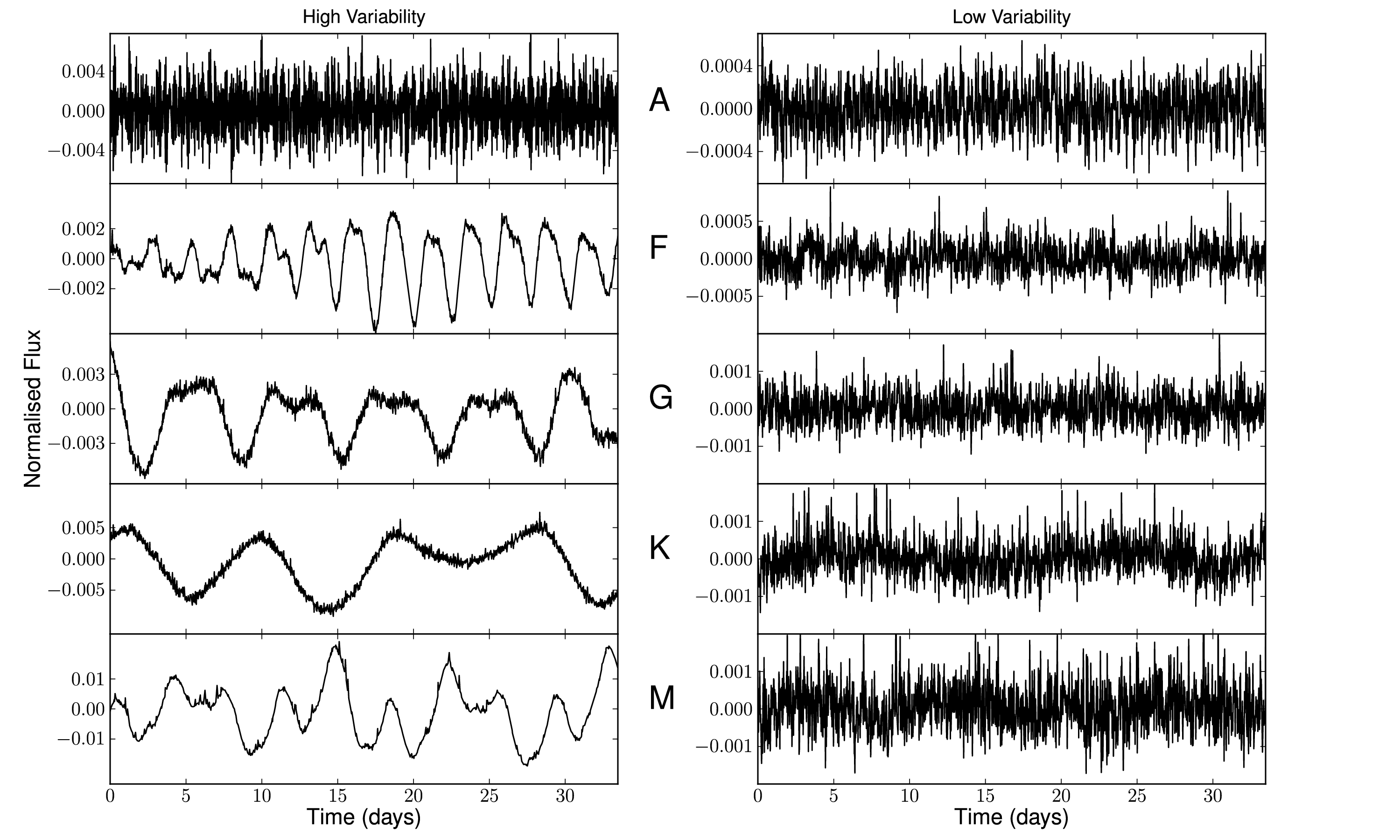}
  \caption{Typical light curves from the high and low variability groups (\emph{left and right respectively}) for each spectral type.}
  \label{fig:lc_examp}
\end{figure*}
	
\begin{figure} 
  \centering
  \includegraphics[width = \linewidth] {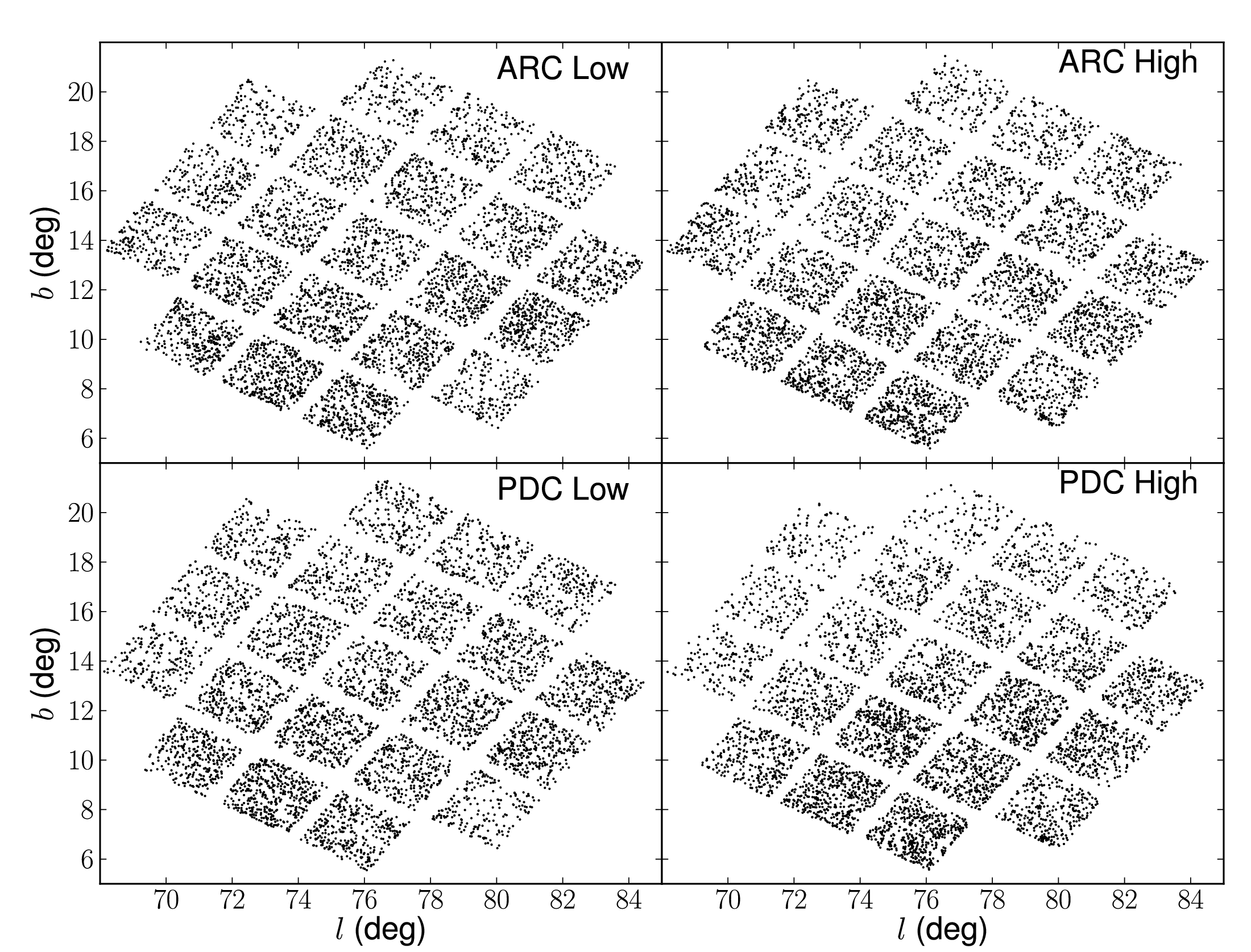}
  \caption{The spatial distribution of low (\emph{left}) and high (\emph{right}) variability stars for ARC (\emph{top}) and PDC (\emph{bottom}). Each has been subject to a random selection of equal numbers of stars per magnitude bin, before a random selection of 6000 for each panel was selected. {\Ared The PDC panels show a slight tendency for the high variability sample to be more concentrated towards the galactic plane, whereas this effect is not significant in the ARC, suggesting it is more effective at removing contamination related systematics.}}
  \label{fig:latlon_dist}
\end{figure}

To examine whether the high and low variability samples belong to different stellar populations, we investigated the possible differences in their spatial distribution and kinematics. Due to dynamical heating of the galactic disk over long timescales, older populations of stars tend to have larger galactic scale heights and larger velocity dispersions than younger populations  \citep[see e.g.][and references therein]{fre02}. We therefore examined the distribution of the two samples in galactic coordinates, which is shown in Figure~\ref{fig:latlon_dist}. 

The division in $R_{\mbox{var}}$ introduces a magnitude bias between the high and low samples, with a greater proportion of the low variability objects at higher magnitudes. To reduce any possible effect of this bias on our comparison tests, we selected high and low variability subsets with approximately equal magnitude distributions. This was done by dividing the stars into 1 mag bins (with the exception of stars brighter than 8th and fainter than 16th in \emph{Kepler} magnitude, which were treated separately), and choosing a random set from the more populated bin to match the number in the smaller one. This process, used for the spatial distribution and proper motion comparison tests, also serves to select an equal number of stars in the high and low variability sets.

When using the PDC data, a difference in spatial distribution of the two subsets is evident, as noted by C11. The low variability stars appear approximately uniformly distributed, whereas the high variability stars are concentrated towards low galactic latitudes. Using the PDC data alone, one could hypothesise that the variable sample may correspond to a younger, thin disk population, and the quiet sample to an older population with larger scale height. C11 hinted at this conclusion but argued that this could also be an artefact of increased crowding at low galactic latitudes, causing higher levels of photometric dispersion.

\begin{figure} 
  \centering
  \includegraphics[width = \linewidth] {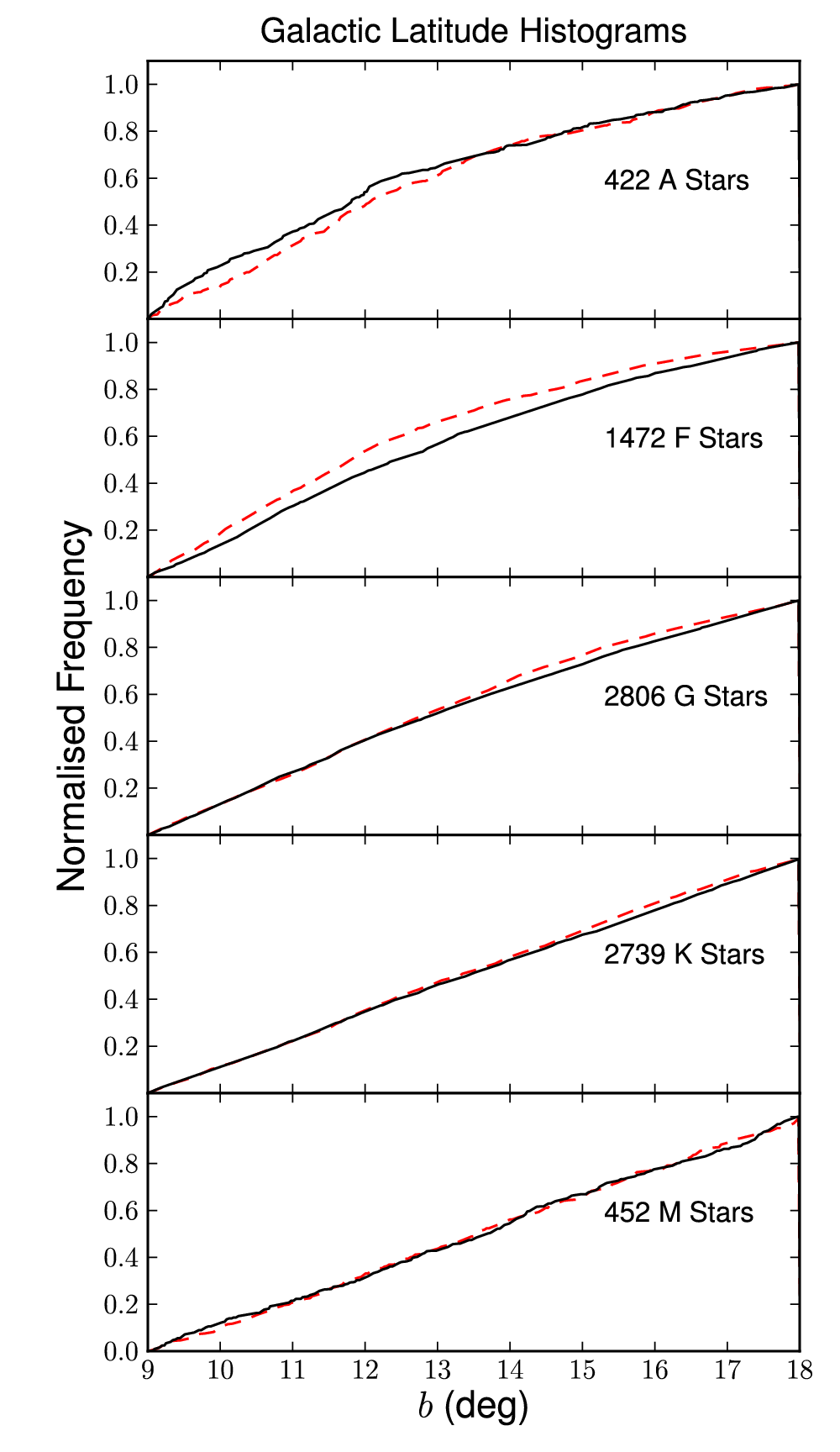}
  \caption{Histogram showing the galactic latitude distribution of low (solid line) and high (dashed line) variability stars for each spectral type. Sample selection ensured the orientation of the \emph{Kepler} field on the plane did not introduce biases. An equal sample of high and low variability stars from each magnitude bin was also selected. {\Ared The increased number of high variability F stars at low galactic latitudes may arise from giant contamination of the sample (see Section~\ref{sec:stelprop}).}}
  \label{fig:glatlon_hist}
\end{figure}

By visually comparing the spatial distribution of the high and low variability samples from the PDC and ARC, it becomes evident that the effect is dependant on the choice of reduction method (Figure~\ref{fig:latlon_dist}).  This suggests that contamination effects were the dominant cause of the apparent higher variability close to the galactic plane, and that the ARC is better at removing the associated systematics. There is still some sign that active F stars are more concentrated at lower latitudes (Figure~\ref{fig:glatlon_hist}), but it is too subtle to draw any strong conclusions. 

The giant contamination in the F stars is higher than other spectral classes because they span the temperature range where the giant branch intersects the main sequence on the HR diagram. This may lead to the apparent concentration of variable F stars at low galactic latitudes, if the variable and luminous giants are visible out to greater distances in the galactic plane, and hence appear with higher density in these regions. 

We also performed a more quantitative two-sample Kolmogorov-Smirnov (KS) test. The results are shown in Table~\ref{table:glat_tab}, and confirm that the galactic latitudes of the low and high variability F stars are very unlikely to be drawn from the same distribution, compared to a high probability for the M stars. The distribution shape parameters, also shown in Table~\ref{table:glat_tab}, display the trends of increasing median galactic latitude with later spectral type and the subtle differences between high and low variability subsets, although these are not sufficient to prove the stars belong to different populations. The spatial distribution plots also serve to show that the correction applied using ARC affects all areas of the CCD equally, despite being implemented on each mod.out separately.

\begin{figure} 
  \centering
  \includegraphics[width = \linewidth] {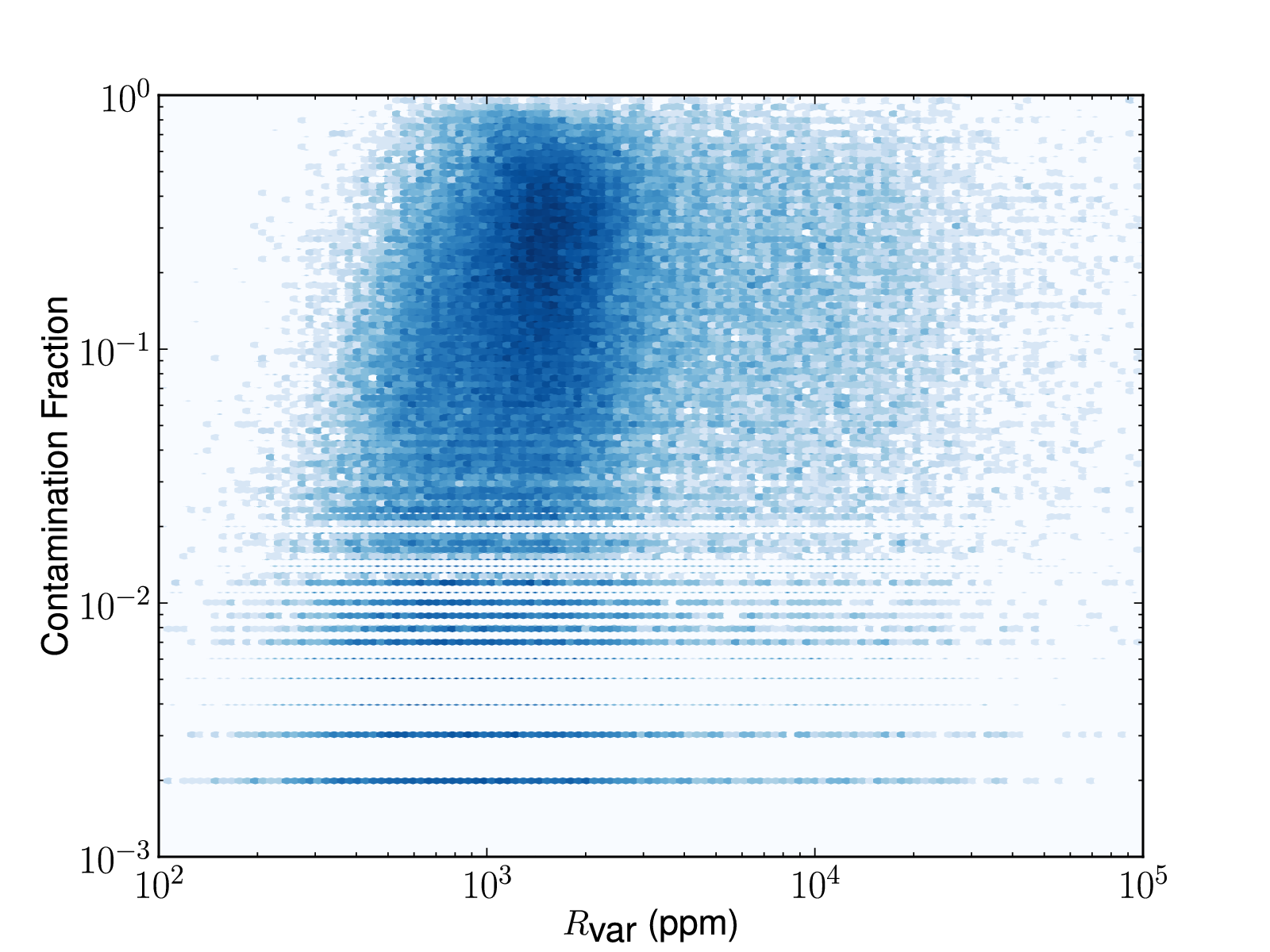}
  \caption{A weak correlation can be seen between contamination fraction and $R_{\mbox{var}}$, for non-zero contamination values.}
  \label{fig:contam_range}
\end{figure}

To examine the effect of crowding, we used the contamination fraction available from the \emph{Kepler} mission archive, which provides an approximate measure of the fraction of flux attributed to the target in a 21 $\times$ 21 pixel aperture. Values range from 0, implying no contamination, to 1, which indicates the flux is essentially all background. There is a weak correlation between contamination fraction and $R_{\mbox{var}}$, shown in Figure~\ref{fig:contam_range}. {\Ared We tested whether placing a constraint on the contamination fraction, using only targets in the range 0.1 to 0.2, altered the appearance of Figure~\ref{fig:latlon_dist}, but it did not.}

\begin{table*}
  \caption{Statistics of galactic latitude ($b$) distributions, including two-sample Kolmogorov-Smirnov tests, median, median absolute deviation (MAD) and skew of the low (L) and high (H) variability samples.}
  \label{table:glat_tab}
  \centering
  \begin{tabular}{lcccccccccc} 
    \hline\hline
    Sample &  $b$ KS &   Med L$b$  & Med H$b$  & MAD L$b$  & MAD H$b$  & Skew L$b$   & Skew H$b$  \\
   \hline 
    A      &   0.016 & 11.68 & 12.00 & 1.76 & 1.57 & 0.70 & 0.54    \\  
    F    &   1.4$\times 10^{-9}$  & 12.52 & 11.79 & 1.92 & 1.70 & 0.35 & 0.59 \\  
    G    & 0.002 & 12.88 & 12.65  & 2.08 & 1.95 & 0.21 & 0.28  \\  
    K     & 0.1  & 13.29  & 13.32 & 2.21 & 2.08 & 0.07 & 0.04  \\  
    M  & 1.0 & 13.65 & 13.58 & 2.21 & 2.08 & -0.004 & -0.016   \\   
   \hline
 \end{tabular}
\end{table*}

\begin{table*}
  \caption{Statistics of proper motion distributions, including two-sample Kolmogorov-Smirnov tests, median, median absolute deviation (MAD) and skew of the low (L) and high (H) variability samples.}
  \label{table:pm_tab}
  \centering
  \begin{tabular}{lcccccccccc} 
    \hline\hline
    Sample &   $pm_{\mbox{total}}$ KS & Med L$pm_{\mbox{total}}$ & Med H$pm_{\mbox{total}}$ & MAD L$pm_{\mbox{total}}$ & MAD H$pm_{\mbox{total}}$ & Skew L$pm_{\mbox{total}}$  & Skew H$pm_{\mbox{total}}$ \\
   \hline 
    A      &     0.61    & 0.0057 & 0.0063 &  0.0032 & 0.0030  & 4.61 & 1.57  \\  
    F    &      0.09   &  0.0077 & 0.0076  & 0.0038 & 0.0033 & 9.57 & 12.66 \\  
    G    & 0.0009   & 0.010  & 0.0091 & 0.0045  & 0.0037 & 12.06 & 4.45 \\  
    K     & 0.002  & 0.013 & 0.011 & 0.0061 & 0.0045 & 8.75 & 9.41 \\  
    M  &   1.5$\times 10^{-6}$  & 0.033 & 0.020 & 0.013 & 0.0091 & 2.41 & 4.48 \\   
   \hline
 \end{tabular}
\end{table*}

\begin{figure} 
  \centering
  \includegraphics[width = \linewidth] {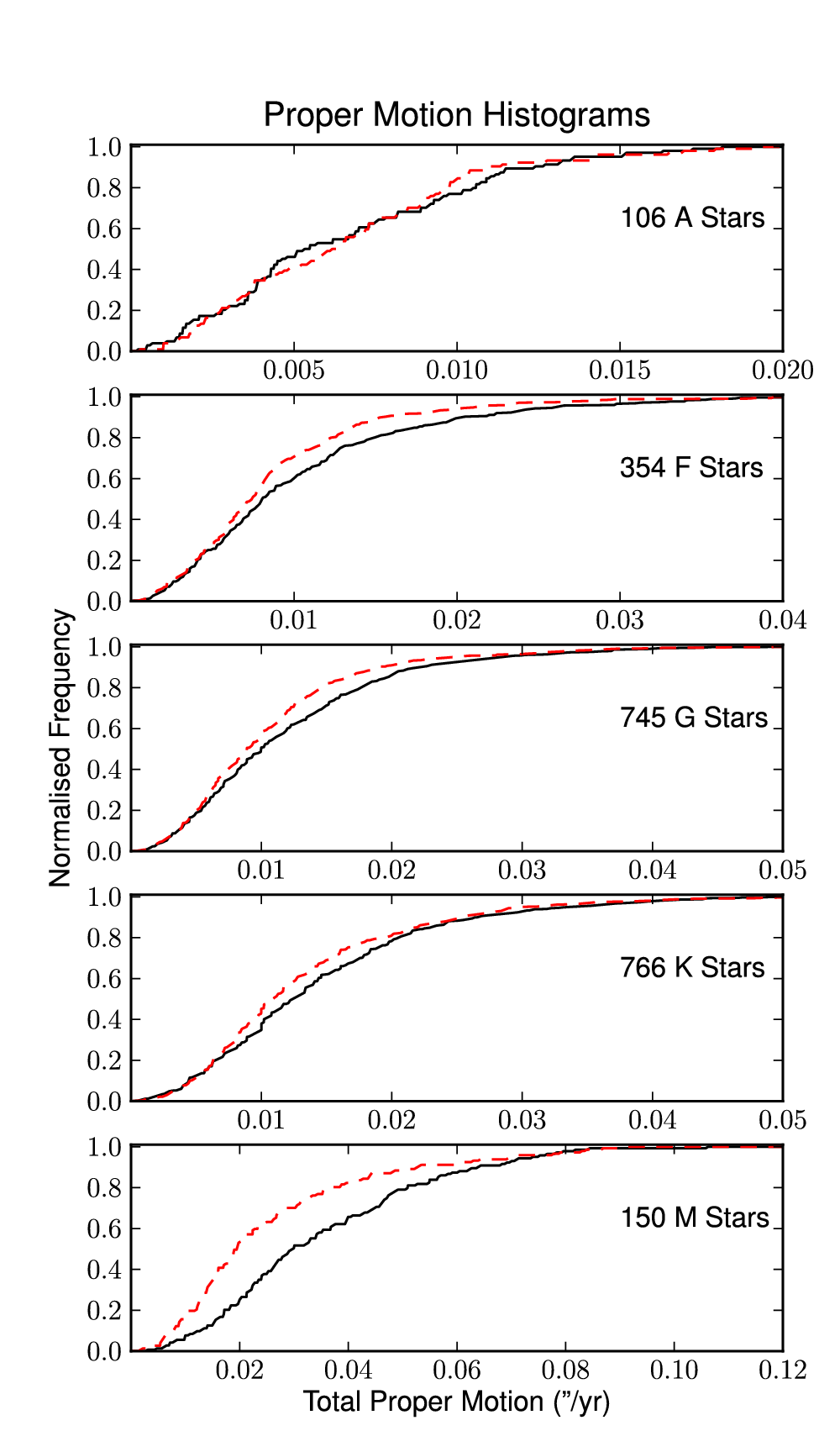}
  \caption{Distribution of proper motion values for the low (solid line) and high (dashed line) variability samples, where proper motion \textgreater\ 0, in a box of even galactic latitude and longitude distribution, with equal numbers of high and low variability stars selected in each magnitude bin. {\Ared The uncertainty on the proper motion values is 0.02''/year but given the high numbers in each sample these results are still significant.}}
  \label{fig:pm_plot}
\end{figure}

A potential indication that the low and high variability subsets belong to different stellar populations can be seen in the proper motion distributions  (Figure~\ref{fig:pm_plot}), for stars with total proper motion greater than zero. It shows a weak but noteworthy trend for the high variability group to have lower proper motion than the low variability group.  We also tested the significance of these differences using two-sample KS tests, and parameterised the shape of the distributions, the results of which can be seen in Table~\ref{table:pm_tab}. The KS test results show that with the exception of A stars, and to some extent F, the proper motion values for the high and low variability groups are very unlikely to be drawn from the same distribution, while the shape parameters highlight more subtle differences in shape and trends between spectral types. This is consistent with the view that higher variability stars are younger and therefore have lower proper motions. This conclusion is dependant on the stars being at the same distance and location on the sky.  In our samples, the magnitude and spatial distributions are approximately equal for the high and low variability subsets, which should satisfy this condition. One caveat applicable to the proper motion distributions is that the giant contamination in the M dwarfs (described by C10) could increase the apparent number of high variability stars with low proper motion. We have also considered potential effects of Malmquist bias in our results, but believe that by selecting equal magnitude distributions and comparing each spectral type separately, any effect introduced should be negligible.

\section{Periodic and Stochastic Variability} 
\label{sec:perstoch}

We investigated the nature of the variability, comparing the low and high variability samples, using a combination of visual examination, the statistics described in Section~\ref{sec:stats_sect}, and periodogram analysis. The typical characteristics of the light curves also vary between spectral types as shown in the examples in Figure~\ref{fig:lc_examp}. Visual examination of the light curves shows that A and F stars contain many pulsators with a large range of amplitudes, while later types show rotational modulation and more stochastic variability. This trend is also apparent in the periodicity and stochasticity tests described in this section.

\subsection{Periodicity}    
\label{sec:per}

For each light curve, we computed a periodogram by least-squares fitting of a sinusoid plus a constant at each trial period. The periodogram is expressed in terms of the statistic
\begin{equation} \label{pgram}
  S = \left(\chi_{0}^{2} - \chi^{2}\right) / \chi_{0}^{2},
\end{equation}
where $\chi^{2}_{0}$ is the reduced chi-squared of the light curve with respect to a constant value, and $\chi^{2}$ is the reduced chi-squared with respect to the best-fit sinusoid. We used 500 logarithmically-spaced periods between 0.01 and 100 days. {\Ared However, only periods between 1 hour (twice the sampling rate) and 16 days (half the time span of the data) were considered valid.} We chose this approach over limiting the initial search space to this range, because this would have led to many spurious detections close to the upper limit. Note that some objects with true periods above 16 days may be detected at harmonics of their true period.  

{\Ared The dearth of stars with periods detected close to the 16 day limit is a bias introduced by the stringent period identification method.  Many objects that would occupy this period range have broad periodogram peaks at around the 16 day limit and the selection method has been designed to only accept periodogram peaks that drop to the median level before reaching the edges of the allowed range.  This prevents false detections of periods at the length of the permitted dataset, and misidentification of the highest periodogram continuum point as a genuine peak. }

\cite{bre98} describes the Bayesian theory of modelling data using a sinusoid plus white noise. In this formalism, our metric can be shown to be
\begin{equation} \label{s_eq}
S = 2C\left(\omega \right)/\left(N\sigma_{\mbox{rms}}^2\right),
\end{equation}

\begin{equation} \label{c_omega}
\mbox{where} (\omega) = \left(R(\omega)^2 + I(\omega)^2 \right) / N,
\end{equation}

\noindent is the squared magnitude of the discrete Fourier transform of the data. $N$ is the number of data points in the light curve, $R(\omega) = \sum_{i=1}^{N}\cos(\omega t_i)$, $I(\omega) = \sum_{i=1}^{N}\sin(\omega t_i)$ and $\sigma_{\mbox{rms}}$ is the rms scatter of the data. We have assumed the data is mean-subtracted.

As discussed in B11, the choice of a threshold for periodicity detection is best made empirically. We selected $S = 0.3$ as an appropriate threshold between clear-cut periodic variability and non-periodic or ambiguous cases. This threshold is slightly lower than the value of $S=0.4$ used in the Monitor project \citep[see e.g.][]{irw06}, which is reasonable given the vastly superior time-sampling of the \emph{Kepler} data. 

This threshold can be compared to a power spectrum cut, since the power spectrum is given by $\mbox{PS}(\omega) = 4C(\omega) / N$ where, following \cite{kje95} we have normalised the power spectra such that a sinusoidal oscillation of amplitude $A$ gives rise to a peak of height $A^2$ in the power spectrum.  Our threshold is then equivalent to a power spectrum threshold of $S_{\mbox{PS}} = 0.6\sigma^2_{\mbox{rms}}$.  By comparison \cite{kje95} give an expression for the noise level in the power spectrum of $\sigma_{\mbox{PS}} = 4 \sigma_{\mbox{rms}} / N$.

In our data, $N \gg 1$, so our threshold is conservative when comparing to white noise. However, white noise is not the dominant factor defining our ability to detect periodicities. To test the appropriateness of our periodicity threshold on data with realistic noise properties, we ran a set of simulations where we injected periodic signals into actual \emph{Kepler} light curves. We randomly  selected 1000 Q1 light curves with low variability ($R_{\rm var} < R_{\rm var,Sun}$) and no significant period ($S < 0.25$), and injected sinusoidal signals into them, with random periods uniformly distributed between 2 and 16 days, and random amplitudes ranging from 0.1 to 10 times the high-frequency noise level (measured as the scatter in the difference between consecutive flux measurements), with a distribution of amplitudes relative to the noise level that was uniform in log. We then stored the best-fit value of $S$ and the corresponding period. We found that $S$ was always $<0.3$ when the best-fit period differed by $>10\%$ from the injected value, suggesting a false detection rate $<10^{-3}$. As one would expect, the missed detection rate is period- and amplitude-dependent: there were only 5 cases in our simulations where the recovered period was within $10\%$ of the injected one but $S$ was $<0.3$, and these were all for $P_{\rm injected} > 12$ days and amplitudes significantly smaller than the noise. However, our simulations did not include a large enough number of these low-amplitude, difficult to detect signals to make a more detailed estimate of the completeness. Since the period search in Q1 data was intended as a preliminary exercise, and the results will become much more reliable when additional data is included, we decided additional simulations were beyond the scope of the present paper.

Using the criteria outlined here, 16\% of the dwarf stars are determined to be periodic, and the fraction for each spectral type is given in Table~\ref{table:var_fract}. We note that these numbers vary slightly from those quoted in B11 and C11. This indicates that our threshold for periodicity is more stringent, designed to detect only genuine periodicities (or strong harmonics) within the limits of the frequency resolution. The periodic fraction will undoubtably increase with longer datasets and more efficient quasi-periodic signal detection methods.

\begin{figure} 
  \centering
  \includegraphics[width = \linewidth] {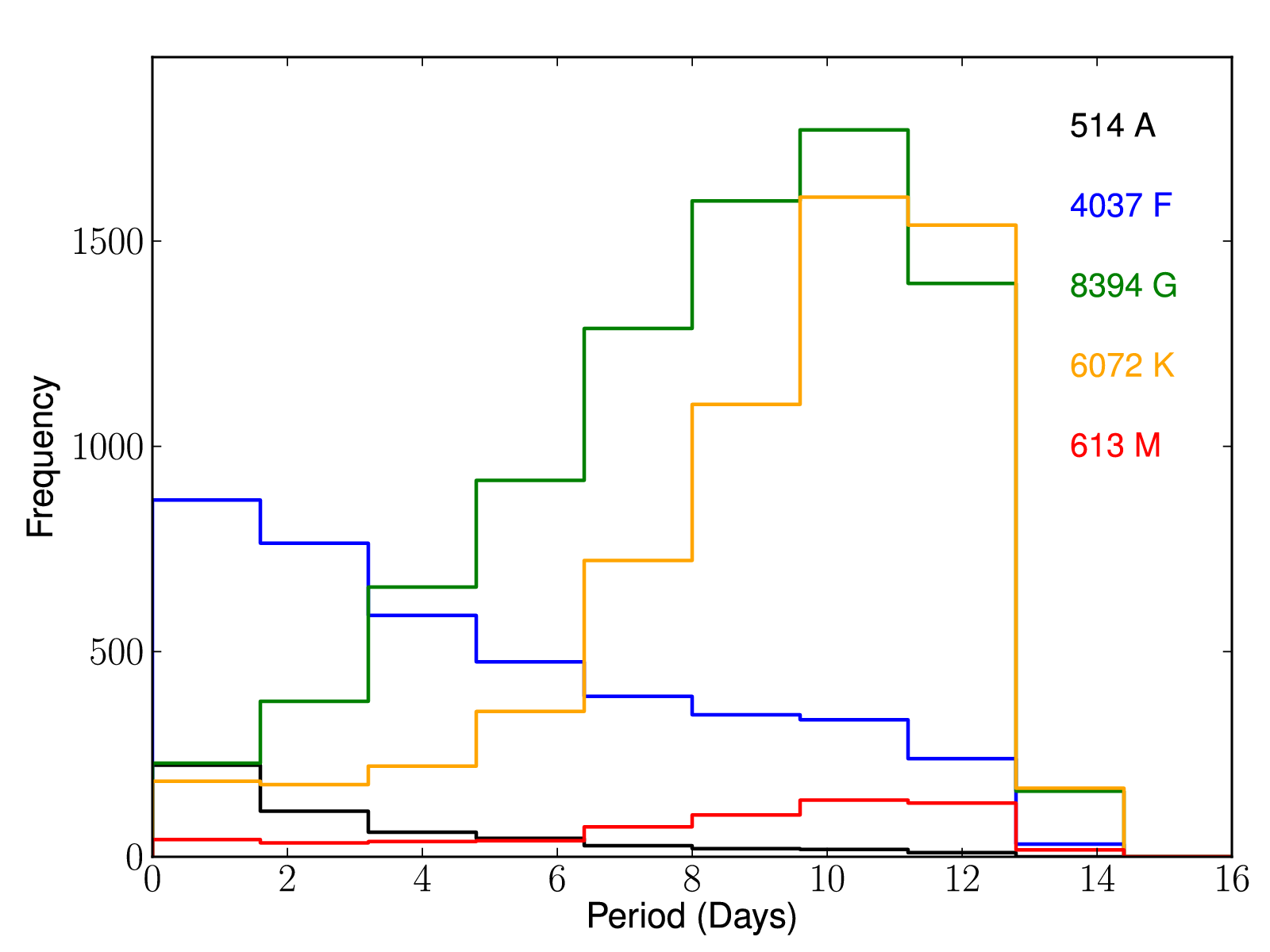}
  \caption{Period distribution for each spectral type, for stars where a significant period between 1hr and 16d has been detected (see Section~\ref{sec:per} for method and caveats), {\Ared showing an increase in period towards later spectral types.}}
  \label{fig:per_plot}
\end{figure}

Histograms of the detected periods for each spectral class are shown Figure~\ref{fig:per_plot}. When interpreting these, one should bear in mind that our period sensitivity is non-uniform, and that the largest peak in the periodogram is not necessarily at the true period but can be at one of its harmonics. The distribution is also truncated close to the maximum period due to the definition of `peak' required for selection. These period distributions should thus be taken as indicative only (the period sensitivity will improve vastly with longer time coverage). Nonetheless, differences between the histograms are obvious, with the typical period clearly increasing towards later spectral types. 

The majority of A stars, and about half of F stars, have very short periods ($<2$ days). These are likely to be pulsators (many of the A and F stars are located in the instability strip), although they could also be unpublished close binaries (where ellipsoidal variations and mutual heating induce sinusoidal variations) or active stars with very short rotation periods. On the other hand, G, K and M stars, along with the rest of the F stars, have significantly longer periods ($\ge 5$ days), as one might expect from rotational modulation of active regions. This distinction is confirmed by the appearance of the light curves (Figure~\ref{fig:lc_examp}). The trend of longer periods towards later spectral types is also visible amongst the late type objects.

\subsection{Degree of Stochasticity}   
\label{sec:stoch}

\begin{figure*} 
  \centering
  \includegraphics[width = \linewidth] {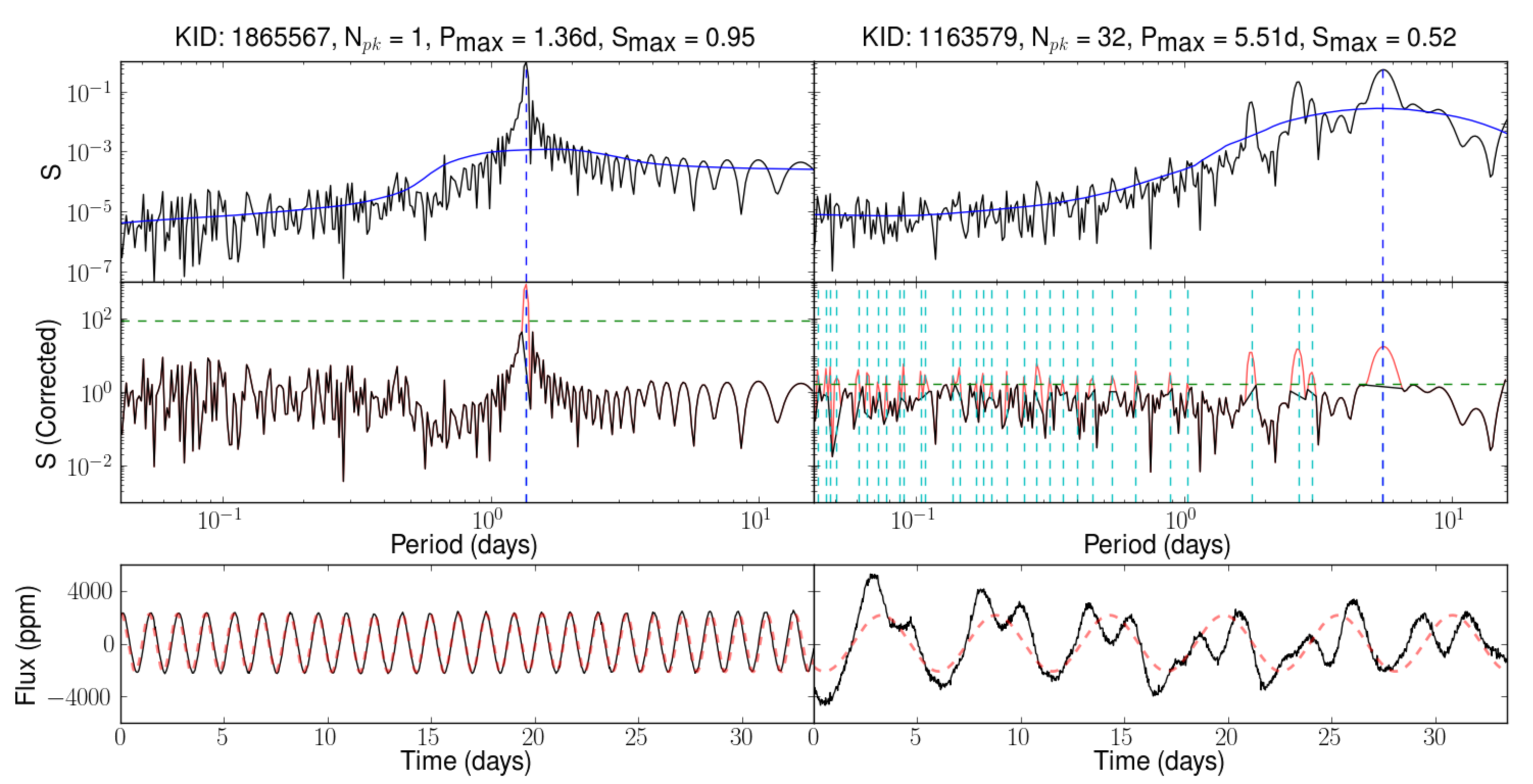}
  \caption{Graphical representation of the calculation of $N_{\rm pk}$ for periodic (\emph{left}) and non-periodic (\emph{right}) stars. The \emph{top panels} show the original power spectrum, smoothed spectrum, and the maximum peak (dashed line). The original power spectrum is divided by the smoothed one to get the corrected spectrum in the \emph{centre panels}. Peaks above 10\% of the maximum value (horizontal dashed line) are then counted.  The bottom panels shows the light curve (solid line) with best fit sinusoid (dashed).}
  \label{fig:npk_calc}
\end{figure*}

A significant difference in the nature of variability across different spectral types is shown by the degree of stochasticity. We performed a simple measure of this based on the number of peaks, $N_{\rm pk}$, in the sine-fitting periodogram which have power greater than 10\% the maximum power. We selected only stars above the periodicity threshold described in Section~\ref{sec:per}.  An example $N_{\rm pk}$ determination is shown in Figure~\ref{fig:npk_calc}.

\begin{figure} 
  \centering
  \includegraphics[width = \linewidth] {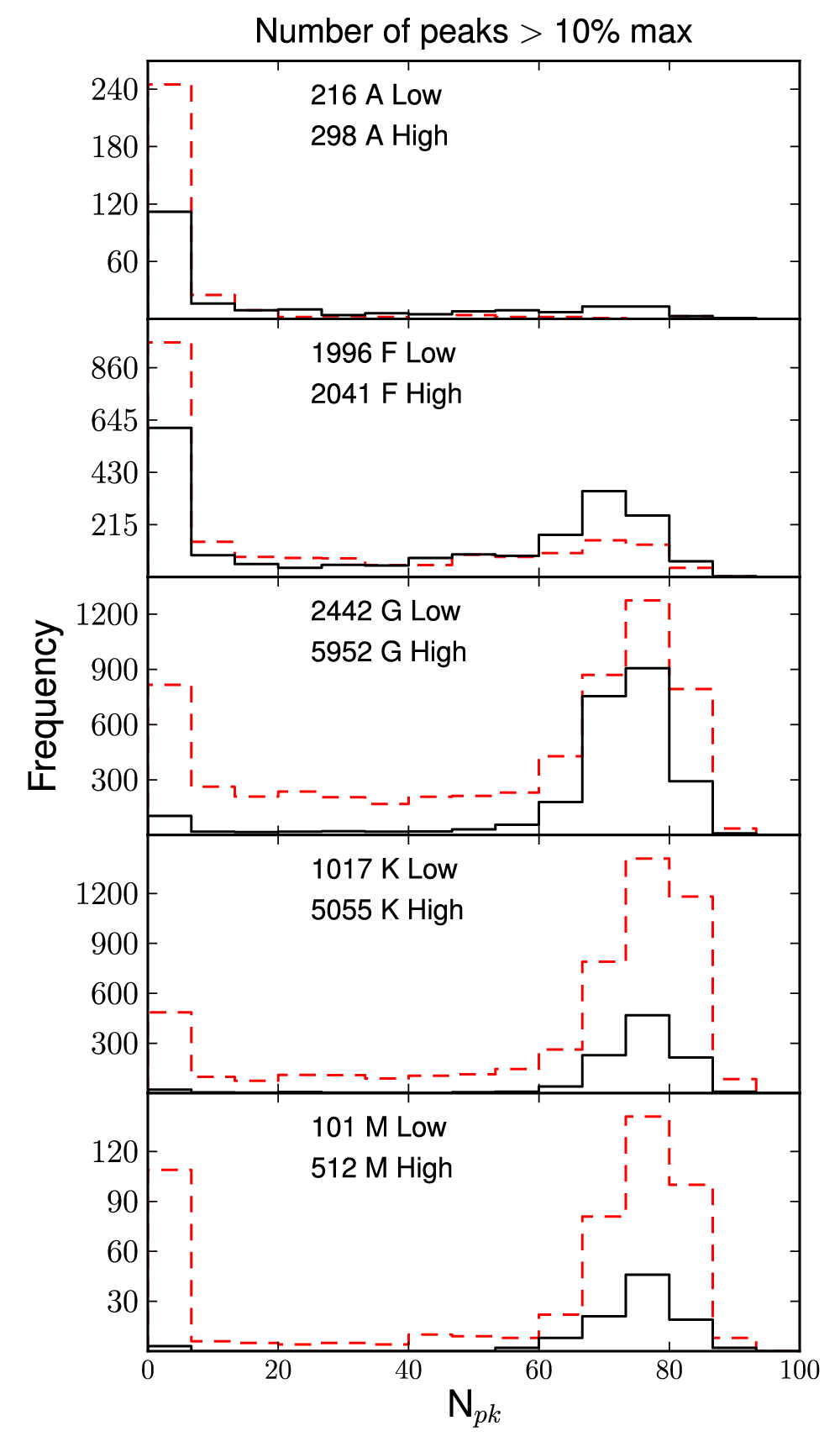}
  \caption{Distribution of the number of peaks with power greater than 10\% of the maximum peak for low (solid line) and high dashed line) variability samples, where $S \ge 0.3$.}
  \label{fig:npk_highlow}
\end{figure}

The histograms showing the $N_{\rm pk}$ distribution for each spectral type are displayed in Figure~\ref{fig:npk_highlow}. The light curves of pulsating stars are dominated by near-sinusoidal variability at one, or a few, clear dominant periodicities, corresponding to low values of $N_{\rm pk}$. Rotational variables are less well modelled by sine-fitting. They typically have a more complex periodogram, with significant power at multiple harmonics of the dominant period \citep[see e.g.][]{boi09} and some stochastic variability (with power at all frequencies, as seen in the case of the Sun, see e.g. \citealt{aig04}). This results in somewhat larger values of $N_{\rm pk}$. {\Ared The high variability group show a tendency towards slightly lower levels of stochasticity than the low variability group, which may arise from the strongly periodic pulsating stars that typically have high amplitude variations.} It is important to note that the $N_{\rm pk}$ statistic is not a quantitive measure of stochasticity, but intended  for use as a comparison between the low and high dispersion samples. 

Alternative metrics for measuring timescale, periodicity and stochasticity were demonstrated by \cite{wal10}, who use time separation between points where the differential light curve crosses zero, and the time separation between changes in the sign of the slope in the light curve. This method, without smoothing of the light curve, is slightly more sensitive to noise than the $N_{\rm pk}$ approach.

\subsection{Harvey Model Fitting}  
\label{sec:harv}

\begin{figure*} 
  \centering
  \includegraphics[width = \linewidth] {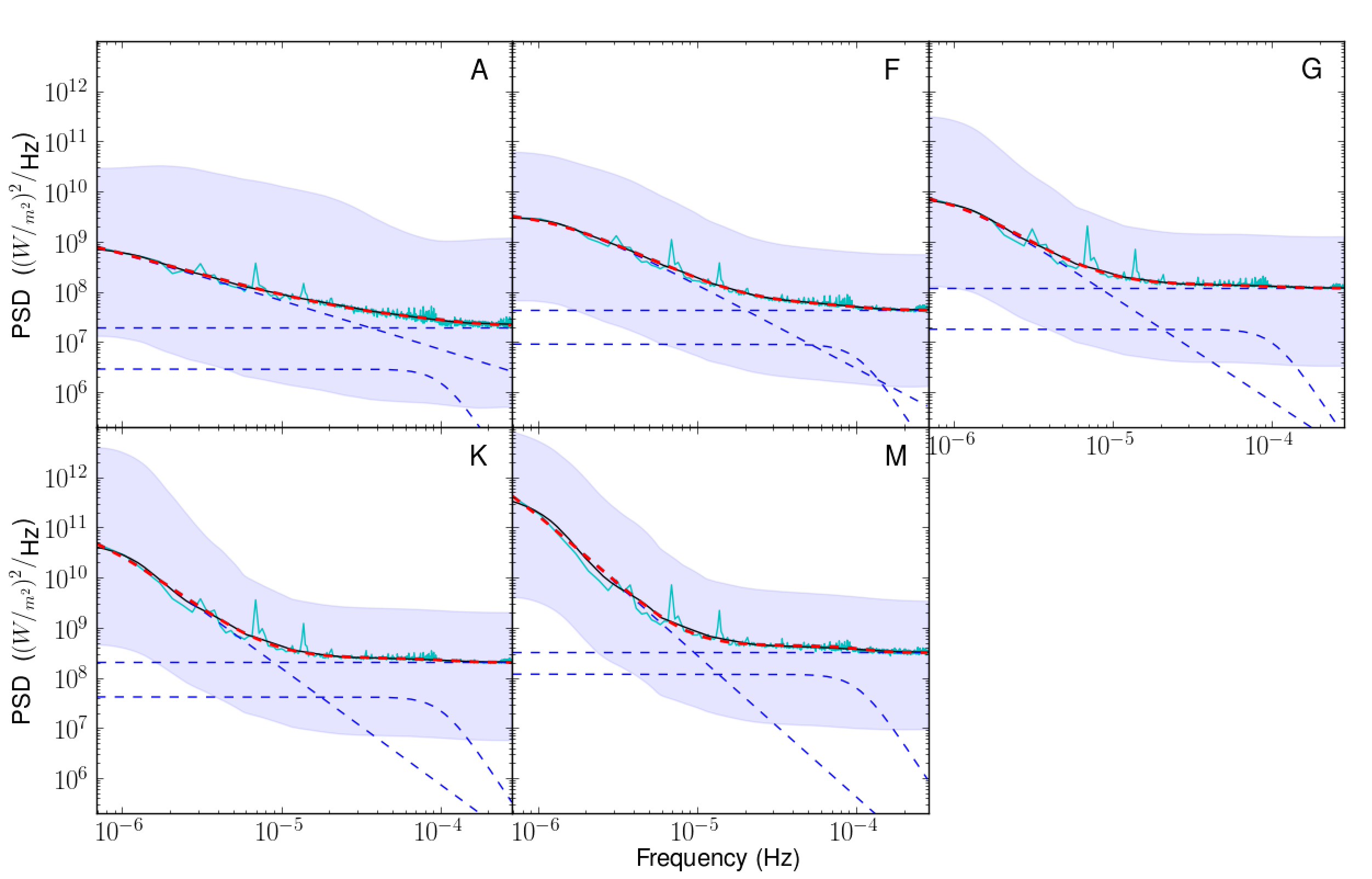}
  \caption{Harvey model fits to the median periodogram for each spectral type. The blue shaded area represents the region between 10-90th percentiles of variability. The cyan line shows the unsmoothed median spectrum, and the fit (red dashed line) is made to the smoothed spectra (black line). The individual components of the fit are shown as blue dashed lines (see text for full explanation).}
  \label{fig:harvey models}
\end{figure*}

We also studied the typical stochastic properties of each spectral class by fitting a model to the average power spectra of each. The Fast Fourier Transform (FFT) of each light curve was computed and the median, 10th and 90th percentile at each frequency were used to create the plots in Figure~\ref{fig:harvey models}. These spectra were smoothed using a nonlinear median boxcar filter to remove the effects introduced by the variable pointing stars (evidence of which can be seen in the peaks of the cyan line in Figure~\ref{fig:harvey models}). 

Autoregressive (AR) models are commonly used to describe stochastic processes and take the form,

\begin{equation} \label{ar}
	x_t = c + \sum_{i=1}^{p}\phi x_{t-1} +  \epsilon_t,
\end{equation}

where $x_t $ and $x_{t-1}$ are data values at time index $t$ and $t$-$1$ respectively, $c$ is a constant, $p$ is the order of the model, $\phi$ are the parameters and $\epsilon$ is white noise. Such a process has a spectral density

\begin{equation} \label{harv}
	P(\nu) = \sum_{i=1}^{N}P_i = \sum_{i=1}^{N} \frac{A_i}{1+ \left(B_i\nu \right)^{C_i}},
\end{equation}

where $\nu$ is frequency, $A_i$ is the amplitude of the $i$th component, $B_i$ is its characteristic timescale, and $C_i$ is the slope of the power law. By considering the power spectra as a sum of $N$ different AR(1) spectra, one can fit a model of this form, first introduced by \cite{har85} and previously used to model stellar spectra \citep[see e.g.][]{aig04, mic09}.

The Harvey model consists of a constant background level, a component to describe the intrinsic power at each frequency, and a second component to describe the extra power at $\sim 8  \times 10^{-5}$ Hz introduced by the onboard motor vibrations.  The timescale and slope of the motor component have been set to the mean value found when the parameters are allowed to vary, because there is no reason they should be different between spectral types. The amplitude for each is allowed to vary because it will have a greater effect for the fainter stars. 

The median power spectrum and fitted powerlaw models describe the typical stochastic background variability of the stars. Clear trends through the spectral classes can be seen in the fit parameters, listed in Table~\ref{table:harv_pars}.  As expected, the background constant increases steadily towards the fainter classes.  The amplitude, timescale and slope of the powerlaw fit also increases towards later spectral types. This may arise from the slower typical rotation periods associated with later type stars, creating large, slowly evolving active regions. The timescale of the A stars does not fit this trend, which is most likely due to the large number of pulsating stars shifting the median power towards slightly longer timescales, and the fact that the frequency resolution is not sufficient to resolve the powerlaw turnoff.

\begin{table}
  \caption{Best fit parameters for the Harvey models shown in Figure~\ref{fig:harvey models}.}
  \label{table:harv_pars}
  \centering
  \begin{tabular}{lccccccc} 
    \hline\hline
    SpT &  Background &  $A_1$  & $B_1$  & $C_1$ & $A_2 $  & $B_2$  & $C_2$  \\
   \hline
    A   &   1.9e7  &    4.1e9     &   1.0e6     &    1.0      &    2.9e6   &   1.6e3  &   4.8       \\    
    F   &   4.3e7  &    4.3e9      &   1.2e5     &    1.7      &   9.1e6   &   1.6e3  &   4.8       \\      
    G   &  1.2e8   &   1.1e10    &    1.6e5    &     2.1     &   1.8e7   &   1.6e3  &   4.8       \\    
    K   &   2.0e8  &    1.2e11    &    2.8e5    &     2.3     &    4.2e7  &   1.6e3  &   4.8       \\    
    M   &   3.2e8  &    1.9e12   &    3.6e5    &     2.8     &   1.2e8   &   1.6e3  &   4.8       \\ 
   \hline
 \end{tabular}
\end{table}

\section{Discussion and Conclusions} 
\label{sec:disc}

We have developed an astrophysically robust systematics correction
method (ARC), which has been shown to remove the significant
systematic trends from the \emph{Kepler} Q1 light curves, while maintaining
intrinsic stellar variability. Figure~\ref{fig:comp_var} shows that
the ARC performs to the same standard as the PDC in reducing the lower
envelope of the measured variability down to the photometric
uncertainty. Crucially, unlike the PDC it does not remove real
astrophysical signatures (predominantly at medium variability levels),
leading to the apparent bimodality in variability level that can be
seen in the PDC panel of Figure~\ref{fig:comp_var}. The correction is
applied individually to each mod.out but the results are shown to be
consistent across the whole field of view
(Figure~\ref{fig:latlon_dist}).

Following B11, we quantified the variability level of each star as the
interval between the 5$^{\rm th}$ and 95$^{\rm th}$ percentile of its
light curve, and we focused our study on the main-sequence stars,
separating them from giants using the $T_{\rm eff}$-dependent $\log g$
cut defined by C10. A division between low and high variability was
made using a magnitude dependent cut in $R_{\mbox{var}}$ at twice the
equivalent variability measurement for the active Sun, resulting in a
split of 80\% low and 20\% high. We examined the relationship between
the nature of the variability and physical properties of the stars,
and compared our work to that of C11, who use the PDC data and B10,11,
who use a polynomial fitting correction method. As suggested by B10
and C11, we see clear evidence for a decrease in variability with
increasing temperature (Figure~\ref{fig:temp_plot}), which cannot be
entirely explained by selection effects or the noise floor at fainter
magnitudes.  This is equivalent to an increase of variability with
increasing $B-V$, as noted by \cite{isa10}, who use $S_{\mbox{HK}}$ as
an indicator of chromospheric activity.

The spatial distribution and proper motion of the stars were used to
look for an indication that the low and high variability samples come
from different stellar populations.  C11 suggest that the high
variability sample belongs to a younger population, existing closer to
the galactic plane than the older low variability sample.  {\Ared We note a weak
correlation between contamination fraction and variability, which
could potentially increase the apparent variability in the more
contaminated regions close to the galactic plane. The proper motion
distributions displayed in Figure~\ref{fig:pm_plot}, indicate that, for the M stars at least, 
more active stars tend to have slightly lower proper motions. This
may suggest they are younger, although distance estimates and radial velocity measurements would be
needed to interpret the proper motion distributions robustly, and it should be noted that
giant contamination could affect this result.}

There may exist interesting further correlations between the
variability level and other parameters listed in the KIC. For example,
we compared the surface gravity and metallicity at different
variability levels, and did find apparently statistical significant
differences (in K-S test terms). However, we chose not to report them
here, because these parameters have larger uncertainties than $T_{\rm
eff}$, and are effected by non-negligible biases as a function of
Galactic position \citep[because of crowding and extinction, see
e.g.][and references therein]{ver11}, which cannot easily be
disentangled from actual differences between populations. Any 
biases affecting the $\log g$ information in the KIC may also
have an indirect effect on the present study because they could affect
the separation between dwarfs and giants. \cite{ver11, bro11}
investigated the reliability of the KIC parameters in detail, and
concluded that the $T_{\rm eff}$ and $\log g$ estimates are reliable
within the stated uncertainties. This implies that any biases should
have only a very minor effect on the C11 dwarf/giant cut which we
used.

One of the most interesting results in this work is the period
distribution for each spectral type, displayed in
Figure~\ref{fig:per_plot}. Although this should be considered a
preliminary test due to the short timespan of the dataset, there is
clear evidence that, for light curves with a clear periodic component,
the period increases towards later spectral types.  The fraction of
periodic stars also varies with spectral class, increasing towards
later types to a maximum in the K stars, before reducing again in M
stars. The relatively high number of stars in which we were able to
identify a periodic or quasi periodic modulation in this dataset alone
suggests that, when further quarters of \emph{Kepler} data are added, it
should be possible to measure periods for a significant fraction of
all \emph{Kepler} stars -- and thus calibrate the evolution of angular
momentum for intermediate and low-mass stars on the main sequence to
an unprecedented level.

We also investigated the stochastic component of the variability in
two different ways. First, we measured the number of periodogram peaks
with power greater than 10\% of the maximum for the stars which pass
our periodicity selection threshold. For all spectral type, the
distribution of this statistic, which we call $N_{\rm pk}$, has a peak
at low values, corresponding to clearly periodic light curves
dominated by a single frequency, or a small number of frequencies,
{\Ared as would be expected for pulsating stars.} However, there is also another peak at high $N_{\rm pk}$,
whose amplitude increases towards later spectral types, reaching a
maximum in the K stars and decreasing slightly in the M stars (as did
the periodicity fraction). This peak corresponds to quasi-periodic
light curves, as expected for rotational variables with evolving
active regions.

We also parameterised the stochastic component of variability was
performed by fitting auto-regressive models (also known as Harvey
models) to the median power spectrum for each spectral type. Because
of the relatively limited frequency resolution (caused by the short
duration of the dataset used in this study), the fitted model
parameters cannot be considered definitive, but they enable a
preliminary inter-comparison. We find that the typical amplitude,
timescale and power-law index all increase towards later types.

This is consistent with our other tests, and supports a broadly
coherent picture of main-sequence variability. The hotter, earlier
spectral classes show a lower variability level on the whole, and the
variables tend to show clearly periodic behaviour on short
time-scales, as expected from pulsations. The shape of their power
spectra suggest that these stars possess smaller active regions that
evolve more quickly.  By contrast, the cooler, later type stars show
larger amplitude variability on longer time-scales, with
quasi-periodic rather than periodic behaviour, and appear to possess
more slowly evolving, larger active regions. The K stars have the
highest fraction of  light curves with a significant periodicity, but
these are also the most complex, with the largest number of
significant frequencies per object.

We are currently working towards a systematics correction suitable for
the Q2 data, which requires treatment for discontinuities and safe
mode effects in conjunction with the ARC method.  Once complete, this
will allow us to perform a similar investigation on a much longer
dataset, providing vastly improved constraints on the periodicity
measurements, and revealing changes in the variability though time
(i.e.\ pulsation and activity cycles). These measurements will improve
with each new release of \emph{Kepler} data. We will also seek to compare our 
variability measurements to other metrics and timescales, for example 
the CDPP on transit timescales calculated by \cite{gil11}. Finally, the results 
presented in this paper also have implications for planet detection and 
followup, in both photometric and radial velocity programs \citep{aig11}, which we 
will attempt to quantify in future work.


\begin{thebibliography}{0}
\expandafter\ifx\csname natexlab\endcsname\relax\def\natexlab#1{#1}\fi

\end{thebibliography}


\begin{thebibliography}{30}
\expandafter\ifx\csname natexlab\endcsname\relax\def\natexlab#1{#1}\fi

\bibitem[{{Aigrain} {et~al.}(2004){Aigrain}, {Favata}, \& {Gilmore}}]{aig04}
{Aigrain}, S., {Favata}, F., \& {Gilmore}, G. 2004, \aap, 414, 1139

\bibitem[{{Aigrain} {et~al.}(2011){Aigrain}, {Pont}, \& {Zucker}}]{aig11}
{Aigrain}, S., {Pont}, F., \& {Zucker}, S. 2011, \mnras, in press

\bibitem[{{Baglin}(2003)}]{bag03}
{Baglin}, A. 2003, Advances in Space Research, 31, 345

\bibitem[{{Basri} {et~al.}(2011){Basri}, {Walkowicz}, {Batalha}, {Gilliland},
  {Jenkins}, {Borucki}, {Koch}, {Caldwell}, {Dupree}, {Latham}, {Marcy},
  {Meibom}, \& {Brown}}]{bas11}
{Basri}, G., {Walkowicz}, L.~M., {Batalha}, N., {et~al.} 2011, \aj, 141, 20

\bibitem[{{Basri} {et~al.}(2010){Basri}, {Walkowicz}, {Batalha}, {Gilliland},
  {Jenkins}, {Borucki}, {Koch}, {Caldwell}, {Dupree}, {Latham}, {Meibom},
  {Howell}, \& {Brown}}]{bas10}
{Basri}, G., {Walkowicz}, L.~M., {Batalha}, N., {et~al.} 2010, \apjl, 713, L155

\bibitem[{{Batalha} {et~al.}(2010){Batalha}, {Borucki}, {Koch}, {Bryson},
  {Haas}, {Brown}, {Caldwell}, {Hall}, {Gilliland}, {Latham}, {Meibom}, \&
  {Monet}}]{bat10}
{Batalha}, N.~M., {Borucki}, W.~J., {Koch}, D.~G., {et~al.} 2010, \apjl, 713,
  L109

\bibitem[{{Boisse} {et~al.}(2009){Boisse}, {Moutou}, {Vidal-Madjar}, {Bouchy},
  {Pont}, {H{\'e}brard}, {Bonfils}, {Croll}, {Delfosse}, {Desort}, {Forveille},
  {Lagrange}, {Loeillet}, {Lovis}, {Matthews}, {Mayor}, {Pepe}, {Perrier},
  {Queloz}, {Rowe}, {Santos}, {S{\'e}gransan}, \& {Udry}}]{boi09}
{Boisse}, I., {Moutou}, C., {Vidal-Madjar}, A., {et~al.} 2009, \aap, 495, 959

\bibitem[{{Borucki} {et~al.}(2010){Borucki}, {Koch}, {Basri}, {Batalha},
  {Brown}, {Caldwell}, {Caldwell}, {Christensen-Dalsgaard}, {Cochran},
  {DeVore}, {Dunham}, {Dupree}, {Gautier}, {Geary}, {Gilliland}, {Gould},
  {Howell}, {Jenkins}, {Kondo}, {Latham}, {Marcy}, {Meibom}, {Kjeldsen},
  {Lissauer}, {Monet}, {Morrison}, {Sasselov}, {Tarter}, {Boss}, {Brownlee},
  {Owen}, {Buzasi}, {Charbonneau}, {Doyle}, {Fortney}, {Ford}, {Holman},
  {Seager}, {Steffen}, {Welsh}, {Rowe}, {Anderson}, {Buchhave}, {Ciardi},
  {Walkowicz}, {Sherry}, {Horch}, {Isaacson}, {Everett}, {Fischer}, {Torres},
  {Johnson}, {Endl}, {MacQueen}, {Bryson}, {Dotson}, {Haas}, {Kolodziejczak},
  {Van Cleve}, {Chandrasekaran}, {Twicken}, {Quintana}, {Clarke}, {Allen},
  {Li}, {Wu}, {Tenenbaum}, {Verner}, {Bruhweiler}, {Barnes}, \& {Prsa}}]{bor10}
{Borucki}, W.~J., {Koch}, D., {Basri}, G., {et~al.} 2010, Science, 327, 977

\bibitem[{{Borucki} {et~al.}(2011){Borucki}, {Koch}, {Basri}, {Batalha},
  {Brown}, {Bryson}, {Caldwell}, {Christensen-Dalsgaard}, {Cochran}, {DeVore},
  {Dunham}, {Gautier}, {Geary}, {Gilliland}, {Gould}, {Howell}, {Jenkins},
  {Latham}, {Lissauer}, {Marcy}, {Rowe}, {Sasselov}, {Boss}, {Charbonneau},
  {Ciardi}, {Doyle}, {Dupree}, {Ford}, {Fortney}, {Holman}, {Seager},
  {Steffen}, {Tarter}, {Welsh}, {Allen}, {Buchhave}, {Christiansen}, {Clarke},
  {Das}, {D{\'e}sert}, {Endl}, {Fabrycky}, {Fressin}, {Haas}, {Horch},
  {Howard}, {Isaacson}, {Kjeldsen}, {Kolodziejczak}, {Kulesa}, {Li}, {Lucas},
  {Machalek}, {McCarthy}, {MacQueen}, {Meibom}, {Miquel}, {Prsa}, {Quinn},
  {Quintana}, {Ragozzine}, {Sherry}, {Shporer}, {Tenenbaum}, {Torres},
  {Twicken}, {Van Cleve}, {Walkowicz}, {Witteborn}, \& {Still}}]{bor11}
{Borucki}, W.~J., {Koch}, D.~G., {Basri}, G., {et~al.} 2011, \apj, 736, 19

\bibitem[{{Bretthorst}(1998)}]{bre98}
{Bretthorst}, G.~L. 1998, Bayesian spectrum analysis and parameter estimation,
  Tech. rep., Lecture Notes in Statistics, Springer-Verlag. Available at
  http://bayes.wustl.edu/

\bibitem[{{Brown} {et~al.}(2011){Brown}, {Latham}, {Everett}, \&
  {Esquerdo}}]{bro11}
{Brown}, T.~M., {Latham}, D.~W., {Everett}, M.~E., \& {Esquerdo}, G.~A. 2011,
  \aj, 142, 112

\bibitem[{{Castelli} \& {Kurucz}(2004)}]{ck04}
{Castelli}, F. \& {Kurucz}, R.~L. 2004, {in press}, arXiv:0405087

\bibitem[{{Ciardi} {et~al.}(2011){Ciardi}, {von Braun}, {Bryden}, {van Eyken},
  {Howell}, {Kane}, {Plavchan}, {Ram{\'{\i}}rez}, \& {Stauffer}}]{cia10}
{Ciardi}, D.~R., {von Braun}, K., {Bryden}, G., {et~al.} 2011, \aj, 141, 108

\bibitem[{{Freeman} \& {Bland-Hawthorn}(2002)}]{fre02}
{Freeman}, K. \& {Bland-Hawthorn}, J. 2002, \araa, 40, 487

\bibitem[{{Fr{\"o}hlich}(2011)}]{fro11}
{Fr{\"o}hlich}, C. 2011, \ssr, 133

\bibitem[{{Gilliland} {et~al.}(2011){Gilliland}, {Chaplin}, {Dunham},
  {Argabright}, {Borucki}, {Basri}, {Bryson}, {Buzasi}, {Caldwell}, {Elsworth},
  {Jenkins}, {Koch}, {Kolodziejczak}, {Miglio}, {van Cleve}, {Walkowicz}, \&
  {Welsh}}]{gil11}
{Gilliland}, R.~L., {Chaplin}, W.~J., {Dunham}, E.~W., {et~al.} 2011, \apjs,
  197, 6

\bibitem[{{Harvey}(1985)}]{har85}
{Harvey}, J.~W. 1985, Future missions in Solar, heliospheric and space plasma
  physics, ESA SP-235

\bibitem[{Huang {et~al.}(1998)Huang, Shen, Long, Wu, Shih, Zheng, Yen, Tung, \&
  Liu}]{hua98}
Huang, N.~E., Shen, Z., Long, S.~R., {et~al.} 1998, Proceedings of the Royal
  Society of London. Series A: Mathematical, Physical and Engineering Sciences,
  454, 903

\bibitem[{{Irwin} {et~al.}(2006){Irwin}, {Aigrain}, {Hodgkin}, {Irwin},
  {Bouvier}, {Clarke}, {Hebb}, \& {Moraux}}]{irw06}
{Irwin}, J., {Aigrain}, S., {Hodgkin}, S., {et~al.} 2006, \mnras, 370, 954

\bibitem[{{Isaacson} \& {Fischer}(2010)}]{isa10}
{Isaacson}, H. \& {Fischer}, D. 2010, \apj, 725, 875

\bibitem[{{Jenkins} {et~al.}(2010){Jenkins}, {Caldwell}, {Chandrasekaran},
  {Twicken}, {Bryson}, {Quintana}, {Clarke}, {Li}, {Allen}, {Tenenbaum}, {Wu},
  {Klaus}, {Middour}, {Cote}, {McCauliff}, {Girouard}, {Gunter}, {Wohler},
  {Sommers}, {Hall}, {Uddin}, {Wu}, {Bhavsar}, {Van Cleve}, {Pletcher},
  {Dotson}, {Haas}, {Gilliland}, {Koch}, \& {Borucki}}]{jen10}
{Jenkins}, J.~M., {Caldwell}, D.~A., {Chandrasekaran}, H., {et~al.} 2010,
  \apjl, 713, L87

\bibitem[{{Kjeldsen} \& {Bedding}(1995)}]{kje95}
{Kjeldsen}, H. \& {Bedding}, T.~R. 1995, \aap, 293, 87

\bibitem[{{Koch} {et~al.}(2010){Koch}, {Borucki}, {Basri}, {Batalha}, {Brown},
  {Caldwell}, {Christensen-Dalsgaard}, {Cochran}, {DeVore}, {Dunham},
  {Gautier}, {Geary}, {Gilliland}, {Gould}, {Jenkins}, {Kondo}, {Latham},
  {Lissauer}, {Marcy}, {Monet}, {Sasselov}, {Boss}, {Brownlee}, {Caldwell},
  {Dupree}, {Howell}, {Kjeldsen}, {Meibom}, {Morrison}, {Owen}, {Reitsema},
  {Tarter}, {Bryson}, {Dotson}, {Gazis}, {Haas}, {Kolodziejczak}, {Rowe}, {Van
  Cleve}, {Allen}, {Chandrasekaran}, {Clarke}, {Li}, {Quintana}, {Tenenbaum},
  {Twicken}, \& {Wu}}]{koc10}
{Koch}, D.~G., {Borucki}, W.~J., {Basri}, G., {et~al.} 2010, \apjl, 713, L79

\bibitem[{{Michel} {et~al.}(2009){Michel}, {Samadi}, {Baudin}, {Barban},
  {Appourchaux}, \& {Auvergne}}]{mic09}
{Michel}, E., {Samadi}, R., {Baudin}, F., {et~al.} 2009, \aap, 495, 979

\bibitem[{{Monet} {et~al.}(2010){Monet}, {Jenkins}, {Dunham}, {Bryson},
  {Gilliland}, {Latham}, {Borucki}, \& {Koch}}]{mon10}
{Monet}, D.~G., {Jenkins}, J.~M., {Dunham}, E.~W., {et~al.} 2010, {in press},
  ArXiv:1001.0305

\bibitem[{{Pont} {et~al.}(2011){Pont}, {Aigrain}, \& {Zucker}}]{pon10}
{Pont}, F., {Aigrain}, S., \& {Zucker}, S. 2011, \mnras, 411, 1953

\bibitem[{{Slawson} {et~al.}(2011){Slawson}, {Prsa}, {Welsh}, {Orosz},
  {Rucker}, {Batalha}, {Doyle}, {Engle}, {Conroy}, {Coughlin}, {Ames Gregg},
  {Fetherolf}, {Short}, {Windmiller}, {Fabrycky}, {Howell}, {Jenkins}, {Uddin},
  {Mullally}, {Seader}, {Thompson}, {Sanderfer}, {Borucki}, \& {Koch}}]{sla11}
{Slawson}, R.~W., {Prsa}, A., {Welsh}, W.~F., {et~al.} 2011, {in press},
  arXiv:1103.1659

\bibitem[{{Udalski} {et~al.}(2008){Udalski}, {Szymanski}, {Soszynski}, \&
  {Poleski}}]{uda08}
{Udalski}, A., {Szymanski}, M.~K., {Soszynski}, I., \& {Poleski}, R. 2008,
  \actaa, 58, 69

\bibitem[{{van Cleve}(2010)}]{van10}
{van Cleve}, J. 2010, \emph{Kepler} Data Release 6 Notes

\bibitem[{{van Cleve} \& {Caldwell}(2009)}]{vcl09}
{van Cleve}, J. \& {Caldwell}, D.~A. 2009, \emph{Kepler} Instrument Handbook, KSCI
  19033-001, Tech. rep., NASA Ames Research Center

\bibitem[{{Verner} {et~al.}(2011){Verner}, {Chaplin}, {Basu}, {Brown},
  {Hekker}, {Huber}, {Karoff}, {Mathur}, {Metcalfe}, {Mosser}, {Quirion},
  {Appourchaux}, {Bedding}, {Bruntt}, {Campante}, {Elsworth}, {Garc{\'{\i}}a},
  {Handberg}, {R{\'e}gulo}, {Roxburgh}, {Stello}, {Christensen-Dalsgaard},
  {Gilliland}, {Kawaler}, {Kjeldsen}, {Allen}, {Clarke}, \& {Girouard}}]{ver11}
{Verner}, G.~A., {Chaplin}, W.~J., {Basu}, S., {et~al.} 2011, \apjl, 738, L28+

\bibitem[{{Walkowicz} \& {Basri}(2010)}]{wal10}
{Walkowicz}, L.~M. \& {Basri}, G. 2010, in IAU Symposium, Vol. 264, IAU
  Symposium, ed. {A.~G.~Kosovichev, A.~H.~Andrei, \& J.-P.~Roelot}, 469--474

\end{thebibliography}
\bibliographystyle{aa}

\begin{acknowledgements}

This research has made use of the NASA/IPAC/NExScI Star and Exoplanet Database, which is operated by the Jet Propulsion Laboratory, California Institute of Technology, under contract with the National Aeronautics and Space Administration. It was supported by a studentship and a standard grant from the Science and Technology Facilities Research Council. The authors wish to thank F.\ Pont, T.\ Lynas-Gray, D.\ Ciardi, G.\ Basri, Bill Chaplin and D.\ Latham for helpful discussions. AM acknowledges support from an STFC studentship and SA from an STFC standard grant. 

\end{acknowledgements}
  
\end{document}